\documentclass[a4paper,fleqn,usenatbib]{mnras}

\usepackage[T1]{fontenc}
\usepackage{ae,aecompl}

\usepackage{graphicx}
\usepackage{amsfonts}
\usepackage{natbib}
\usepackage{amssymb, amsmath, amsbsy}
\usepackage[dvips]{color}

\defcitealias{bs16}{BS16}
\defcitealias{sb15}{SB15}
\defcitealias{cm77}{CM77}

\newcommand{\kms}{\,km\,s$^{-1}$} 
\newcommand{\msolar}{M_\odot} 
\newcommand{\kelvin}{\,\mathrm{K}}
\newcommand{\cm}{\mathrm{cm}}
\newcommand{\parsec}{\,\mathrm{pc}}
\newcommand{\kB}{k_\mathrm{B}}
\newcommand{\mh}{m_\mathrm{H}}

\newcommand{\fig}{Figure \ref}
\newcommand{\eqn}{Equation \ref}
\newcommand{\tab}{Table \ref}
\newcommand{\sect}{Section \ref}
\newcommand{\bq}{\begin{equation}}
\newcommand{\eq}{\end{equation}}

\newcommand{\qclass}{q_\mathrm{class}}
\newcommand{\qsat}{q_\mathrm{sat}}
\newcommand{\machsat}{\mathcal{M}_\mathrm{sat}}
\newcommand{\machhot}{\mathcal{M}_\mathrm{h}}
\newcommand{\qstar}{q_*}
\newcommand{\sigmastar}{\sigma_*}

\newcommand{\mach}{\mathcal{M}}
\newcommand{\tcc}{\tau_\mathrm{cc}}
\newcommand{\tdiff}{\tau_\mathrm{diff}}
\newcommand{\tmix}{\tau_\mathrm{mix}}
\newcommand{\tsc}{\tau_\mathrm{sc}}
\newcommand{\qs}{\hat{q}_\mathrm{s}}
\newcommand{\fS}{f_\mathrm{s}}
\newcommand{\fr}{f_\mathrm{r}}
\newcommand{\mlra}{\dot{\hat{m}}_\mathrm{A}}
\newcommand{\vrel}{v_\mathrm{rel}}
\newcommand{\mfp}{\lambda_\mathrm{mfp}}
\newcommand{\kappahot}{\kappa_\mathrm{h}}

\newcommand{\mcloud}{M_\mathrm{c}}
\newcommand{\rcloud}{R_\mathrm{c}}
\newcommand{\lcloud}{L_\mathrm{c}}
\newcommand{\pii}{P_\mathrm{II}}
\newcommand{\pevap}{P_\mathrm{ev}}
\newcommand{\csone}{c_\mathrm{s,1}}
\newcommand{\Thot}{T_\mathrm{h}}
\newcommand{\chot}{c_\mathrm{h}}
\newcommand{\nhot}{n_\mathrm{h}}
\newcommand{\rhohot}{\rho_\mathrm{h}}
\newcommand{\ncloud}{n_\mathrm{c}}
\newcommand{\Ncloud}{N_\mathrm{c}}
\newcommand{\pram}{P_\mathrm{ram}}
\newcommand{\fram}{f_\mathrm{ram}}
\newcommand{\vshock}{v_{s}}
\newcommand{\etas}{\eta_\mathrm{s}}
\newcommand{\taus}{\tau_\mathrm{s}}
\newcommand{\betas}{\beta_\mathrm{s}}
\newcommand{\xs}{x_\mathrm{s}}
\newcommand{\tninty}{t_\mathrm{90}}
\newcommand{\tseventyfive}{t_\mathrm{75}}
\newcommand{\tfifty}{t_\mathrm{50}}
\newcommand{\ttwentyfive}{t_\mathrm{25}}
\newcommand{\vexp}{v_\mathrm{exp}}
\newcommand{\cscloud}{c_\mathrm{c}}
\newcommand{\rhocloud}{\rho_\mathrm{c}}
\newcommand{\Tcloud}{T_\mathrm{c}}
\newcommand{\pcloud}{P_\mathrm{c}}
\newcommand{\sigmacloud}{\sigma_\mathrm{c}}
\newcommand{\tkh}{\tau_\mathrm{KH}}
\newcommand{\lkh}{\lambda_\mathrm{KH}}
\newcommand{\fkh}{f_\mathrm{KH}}
\newcommand{\Tii}{T_\mathrm{II}}
\newcommand{\rii}{r_\mathrm{II}}
\newcommand{\nii}{n_\mathrm{II}}
\newcommand{\vii}{u_\mathrm{II}}
\newcommand{\rhoii}{\rho_\mathrm{II}}
\newcommand{\cii}{c_\mathrm{II}}
\newcommand{\facm}{f_\mathrm{m}}
\newcommand{\Lf}{L_\mathrm{F}}
\newcommand{\Teq}{T_\mathrm{eq}}

\newcommand{\midx}{m_\mathrm{i}}
\newcommand{\Nidx}{N_\mathrm{i}}

\newcommand{\figpath}{./figures}

\title[Introduction of PhEW]{A New Model For Including Galactic Winds in Simulations of Galaxy Formation I: Introducing the Physically Evolved Winds (PhEW) Model}

\author[S. Huang et al.]{
Shuiyao Huang$^{1}$\thanks{E-mail:shuiyao@astro.umass.edu},
Neal Katz$^{1}$, Evan Scannapieco$^{2}$, J'Neil Cottle$^{2}$\newauthor
Romeel Dav\'e$^{3,4,5}$, David H. Weinberg$^{6}$, Molly S. Peeples$^{7,8}$ \& Marcus Br\"uggen$^{9}$
\\$^1$ Astronomy Department, University of Massachusetts, Amherst, MA 01003,
USA
\\$^2$ School of Earth and Space Exploration, Arizona State University, P.O. Box 871404, AZ 85287-1404, USA 
\\$^3$ Institute for Astronomy, Royal Observatory, University of Edinburgh, Edinburgh EH9, 3HJ, UK 
\\$^4$ University of the Western Cape, Bellville, Cape Town 7535, South Africa 
\\$^5$ South African Astronomical Observatories, Observatory, Cape Town 7925, South Africa 
\\$^6$ Astronomy Department and CCAPP, Ohio State University, Columbus, OH 43210, USA 
\\$^7$ Space Telescope Science Institute, 3700 San Martin Drive, Baltimore, MD 21218 
\\$^{8}$ Department of Physics \& Astronomy, Johns Hopkins University, 3400 N. Charles Street, Baltimore, MD 21218 
\\$^{9}$ Hamburger Sternwarte, Universit\"at of Hamburg, Gojenbergsweg 112, D-21029, Hamburg, Germany
}

\date{Accepted 0000 October 00. Received 0000 October 00; in original form 0000 October 00}

\pubyear{2019}

\begin{document}
\label{firstpage}
\pagerange{\pageref{firstpage}--\pageref{lastpage}} \pubyear{0000}
\maketitle


\begin{abstract}
The propagation and evolution of cold galactic winds in galactic haloes is crucial 
to galaxy formation models. However, modelling of this process in hydrodynamic simulations 
of galaxy formation is over-simplified owing to a lack of numerical resolution and 
often neglects critical physical processes such as hydrodynamic instabilities 
and thermal conduction.
We propose an analytic model, Physically Evolved Winds (PhEW), that calculates the 
evolution of individual clouds moving supersonically through a uniform ambient medium. 
Our model reproduces predictions from very high resolution cloud-crushing simulations 
that include isotropic thermal conduction over a wide range of physical conditions. 
We discuss the implementation of this model into cosmological hydrodynamic simulations 
of galaxy formation as a sub-grid prescription to model galactic winds more robustly both 
physically and numerically.
\end{abstract}

\section{Introduction}
Many lines of evidence imply that galactic winds are a critical element
of the physics of galaxy formation.  Most directly, observations reveal
ubiquitous outflows from star-forming galaxies at $z\sim 2$
\citep{steidel10} and from starburst or post-starburst galaxies
at low redshift \citep{veilleux20}.
Semi-analytic models and hydrodynamic cosmological simulations
that do not incorporate strong outflows predict galaxies that
are too massive and too metal-rich
(e.g., \citealt{white91,benson03}).
UV absorption studies demonstrate the existence of a cool
($T \sim 10^4\,{\rm K}$), enriched circumgalactic medium (CGM)
with a mass and metal content comparable to or even exceeding that
of the galaxy's stellar component
(e.g. \citealt{tumlinson11,werk14,peeples14,tumlinson17}).
X-ray studies reveal a hot, metal-enriched CGM around elliptical galaxies
\citep[e.g.][]{anderson13}, some massive spirals (e.g. \citealt{bogdan13, bogdan17}), 
and the Milky Way (e.g. \citealt{gupta12, gupta17}).
Hydrodynamic simulations play a crucial role in interpreting these
observations, but the physical processes
that govern the launch and propagation of winds are uncertain and
may occur on scales well below the resolution limit of the 
simulations.  In this paper we describe a ``sub-grid'' approach to
modelling wind propagation, one that adopts a phenomenological 
description of cold cloud evolution informed by high resolution
numerical studies.

Many mechanisms have been proposed for launching galactic winds,
including radiation pressure from young stars, energy and momentum
injection from stellar winds and supernovae, and cosmic ray pressure
gradients.  Different mechanisms may dominate in different situations,
and in some cases the combination of two or more mechanisms
may be crucial \citep{hopkins12}.  In high mass galaxies,
observational and theoretical evidence suggests that feedback from
accreting supermassive black holes (AGN feedback) becomes the
dominant driver of outflows.  Very high resolution simulations,
some from cosmological initial conditions, others of isolated disks 
or sections of the interstellar medium (ISM), are beginning to
provide insights into the ways that these mechanisms launch outflows
(e.g., \citealt{hopkins12, girichidis16, tanner16, fielding17, kim17, li17, schneider18}). 
Observations frequently reveal the 
co-existence of molecular gas, neutral atomic gas, cool ionised gas, 
and hot X-ray emitting gas in the same outflows
\citep{veilleux20}, with the cold and cool phases often
dominating the total mass.  Accurately modelling the interactions
among these multiple phases is a critical challenge.  The acceleration
of large amounts of cold/cool gas to highly supersonic velocities is
a particular puzzle, with possible mechanisms including radiation
pressure on cold clouds \citep{murray05},
entrainment of cold gas in a hot wind \citep{sb15,schneider17},
the formation of the cold/cool phases out of the hot flow by
radiative cooling \citep{thompson16,schneider18}, 
or many mechanisms combined \citep{yu20}.

Our focus in this paper is not the wind launch process itself but the 
evolution of winds after ejection from the galaxy and their interaction
with the CGM.  Most hydrodynamic simulations of cosmological volumes ---
tens of Mpc on a side, containing many galaxies --- adopt a phenomenological
model in which wind particles are launched stochastically from each 
star-forming galaxy at rates and velocities motivated by analytic
models or by pressure gradients induced with tuned prescriptions of
energy or momentum injection.  Examples include our own group's
simulations (e.g., \citealt{oppenheimer06,dave13,simba,huang20a}) and the 
Illustris \citep{vogelsberger13}
and Illustris TNG \citep{pillepich18a} simulations.  
Other groups add the feedback energy as thermal energy, e.g. \citet{stinson06}, 
EAGLE \citep{schaye15} and FIRE \citep{hopkins12},
and allow the winds to develop as a result.
Cosmological volume simulations enable statistical comparisons
to the observed evolution of galaxy masses, sizes, star formation
rates, and gas content
(e.g., \citealt{oppenheimer10, pillepich18b, simba}), and they have played
an essential role in interpreting UV absorption observations of the CGM
(e.g., \citealt{ford13, ford16, nelson18}).
However, given the potential sensitivity of predictions to physical
processes in the CGM below the resolution limit of the simulations,
independent of how the winds are generated in the simulations,
it is still unclear which empirical successes of the simulations are
true successes and which failures are true failures.
Simulations that deliberately amplify resolution in the CGM offer
one route to examining the impact of resolution on predictions
showing that some quantities can be significantly affected
(e.g., \citealt{hummels19,peeples19,vandevoort19}).
Another approach to increase the resolution is to use ``zoom'' simulations to
model one galaxy at a time 
\citep[e.g.][]{governato07, hopkins14, wang15, grand17}.
However, even these simulations do not resolve the tens of pc-scale
structures suggested by some theoretical models of thermal
instability \citep{mccourt18, mandelker19} and by
estimates of cool-phase cloud sizes inferred from measured
column densities and derived number densities
(e.g., \citealt{pieri14, crighton15, stern16}).  
Furthermore, with
current computational capabilities it is infeasible to maintain
even kpc-scale resolution throughout volumes that are tens of Mpc
on a side, and standard Lagrangian or mesh refinement schemes will
not automatically resolve the regions of the CGM where gas phases
interact.
Physical processes in addition to radiative hydrodynamics, 
such as thermal conduction, viscosity and magnetic fields, may also
have significant effects on cloud 
evolution \citep{marcolini05, orlando05, vieser07b, mccourt15, bs16, 
armillotta16, armillotta17, li19}, 
but they are rarely incorporated self-consistently in cosmological simulations.

In the approach proposed here, we eject wind particles as in previous 
simulations but follow their evolution and interaction with the ambient
CGM using an analytic sub-grid model. We model the gas in each wind
particle as a collection of cold clouds, and we calculate the exchange of
mass, momentum, energy, and metals between these clouds and the surrounding
CGM gas.  A wind particle loses mass as it evolves, and it is dissolved
when its mass falls below some threshold, or when its velocity and physical
properties sufficiently resemble the surrounding gas, or when it rejoins
a galaxy and contributes its remaining mass and metals to the ISM.
This general method can be implemented in cosmological simulations that
use smoothed particle hydrodynamics (SPH) or Eulerian or Lagrangian
mesh codes.  In future work we will present results from implementing
this Phenomenologically Evolved Wind (PhEW) model in cosmological
simulations with the GIZMO hydrodynamics code \citep{hopkins16},
employing the star formation and feedback recipes described by
\citet{simba} and the wind launch prescriptions described by
\citet{huang20a}, which are themselves tuned to reproduce outflows
in the FIRE simulations \citep{muratov15}.  In this paper we present
the wind model itself.

This model is based on results from very high resolution simulations of
the ``cloud-crushing problem,'' which examine the evolution of an 
individual cold clouds moving supersonically relative to an ambient,
hotter flow (see \citealt{banda-barragan19} for a recent compilation of
such simulations).  We concentrate in particular on the cloud-crushing
simulations of \citet[][hereafter \citetalias{sb15}]{sb15}, 
which do not include thermal conduction,
and the simulations of \citet[][hereafter \citetalias{bs16}]{bs16}, which do.  
These simulations
model idealised situations with mass and spatial resolutions far higher
than that achieved in any galaxy formation or cosmological hydrodynamic
simulations.  PhEW provides a method to transfer the lessons from these
high resolution studies to a cosmological context.  This method 
necessarily introduces new free parameters, the most important being
the individual cold cloud mass and the strength of thermal conduction.
However, traditional implementations of galactic winds in cosmological
simulations implicitly introduce a non-physical ``sub-grid'' model that
is governed by the numerics of the interaction between wind particles and
the ambient gas with very different physical properties.  
The effects of this non-physical model (e.g., the degree to which cold
gas remains cold) may be sensitive to the numerical resolution.
PhEW replaces these numerically governed interactions with a model
that is physical, approximate, makes specified and controllable
assumptions, and should be less sensitive to numerical resolution.

The paper is organised as follows. In \sect{sec:ccproblem} we describe the 
set-up of the cloud-crushing problem and the various physical processes involved
and also introduce the cloud-crushing simulations \citepalias{bs16} that we use
to develop the analytic model. In \sect{sec:physics} we discuss different 
physical regimes of the problem and the dominant physics. In \sect{sec:phew} we 
describe our analytic model and provide a detailed calculation of how physical properties 
of the cloud evolve with time. In \sect{sec:assumptions} we summarise the key assumptions 
and approximations in the analytic model and discuss where they might break down. 
In \sect{sec:tests} we compare our model predictions to simulation results from 
\citepalias{bs16}. In \sect{sec:discussion} we summarise the main results from the 
paper, describe how to implement the analytic model in cosmological simulations 
that use various hydrodynamic methods and discuss the implications of this model 
in galaxy formation.


\section{The Cloud-Crushing Problem}
\label{sec:ccproblem}
\begin{figure*}
  \includegraphics[width=1.75\columnwidth]{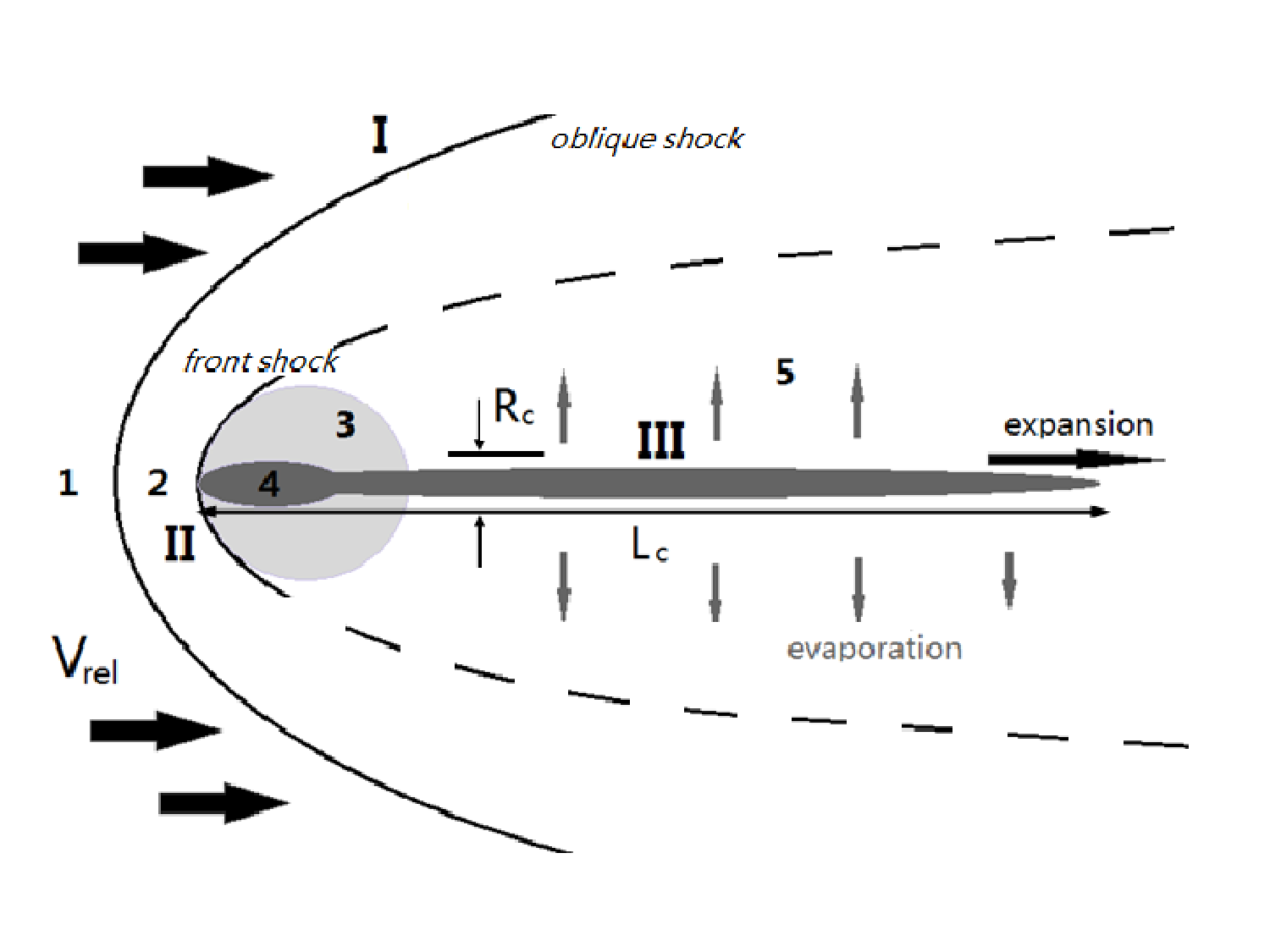}
  \centering
  \caption{Cloud crushing problem set up. See text for description. Here region 
2 represents only the gas that surrounds the head of the cloud. The transition region 
5 between the oblique shock boundary I and the cloud boundary III appears broader 
than is likely in reality. The fluid properties within this region are determined 
by the combined effects of the oblique shock and the evaporative flow.}
  \label{fig:cartoon}
\end{figure*}

We set up the cloud-crushing problem as illustrated in \fig{fig:cartoon}. An initially 
uniform, spherical cloud of mass $\mcloud$, density $\rho_3$ and temperature $T_3$ 
is placed in an ambient medium of uniform density $\rho_1$ and temperature $T_1$ 
with a relative velocity $\vrel$. The initial density contrast between the cloud 
and the ambient medium is $\chi_0 \equiv \rho_3/\rho_1$. 

At the beginning, the cloud is in thermal pressure equilibrium with the surrounding 
medium so that $\rho_3T_3 = \rho_1T_1$. We let the cloud move relative to the ambient 
at a velocity $\vrel$. Here we only study the cases where the cloud is moving supersonically 
as is typical in wind-CGM interactions, i.e., $\mach_1 \equiv \vrel/\csone > 1$, 
where $\csone$ is the isothermal sound speed of the ambient medium. The discontinuity 
in front of the cloud separates into a \textit{bow shock} (region 2) that moves 
into the ambient medium and a \textit{cloud shock} that advances into the cloud. 
We note the surface at the front of the bow shock with roman numeral I\footnote{The 
\textit{oblique shock} on the sides of the cloud is weaker than the \textit{front 
shock} but we do not distinguish them here and use the same notation for the entire 
interface.} and the surface at the contact discontinuity with roman numeral II. 
We define the cloud-crushing time-scale as $\tcc \equiv \chi_0^{1/2}(\rcloud/\vrel)$ 
\citepalias{sb15}, where $\rcloud$ is the initial radius of the cloud. It takes 
$\sim\tcc$ for the cloud shock to sweep through the cloud, crushing it into a much higher 
density $\rho_4$ and pressure $P_4$ that is comparable to the pressure at the stagnation 
point $\pii$. After the cloud shock, the cloud will re-expand preferentially in 
the under-pressured downstream direction. The ambient flow between the shock front 
I and the boundary of the cloud (III) is continuous and obeys Bernoulli's equations.

In cloud-crushing simulations \textbf{without thermal conduction}, the clouds are 
vulnerable to hydrodynamic instabilities. For example, in many cases, perturbations 
grow at the cloud boundary III owing to the Kelvin-Helmholtz Instability (KHI), which eventually 
leads to the fragmentation and disruption of the cloud within a few $\tcc$. Even 
in high Mach flows where the KHI tends to be suppressed, the cloud hardly survives 
beyond 30 $\tcc$ \citepalias{sb15}. 

Cloud-crushing simulations \textbf{with thermal conduction} suggest that efficient 
thermal conduction significantly affects cloud evolution \citep{orlando05, vieser07b, 
bs16, armillotta16, armillotta17, li19}: 
First, the cloud evaporates when thermal conduction is sufficiently strong. In many 
such simulations, cloud evaporation is the leading cause of mass loss. Rapid evaporation 
sometimes destroys the cloud much sooner than without conduction. 
Second, the evaporated material streaming away from the cloud creates a conduction 
zone where the pressure $\pevap$ at the cloud surface III is larger than the thermal 
pressure \citep[][hereafter \citetalias{cm77}]{cm77}.
This helps to confine the cloud and prevent it from 
fragmentation caused by KHI. In \citetalias{bs16}, the clouds often display a needle-like 
morphology as illustrated in \fig{fig:cartoon}, with radius $\rcloud$ and length 
$\lcloud$, instead of breaking up into small clumps, which occurs when there is 
no conduction \citepalias{sb15}.
Third, the deceleration rate of the cloud from ram pressure $\pii$ is reduced because 
its cross section ($\pi\rcloud^2$) shrinks owing to the additional 
vapour pressure. 
Fourth, the jump conditions at the bow shock must be modified from the Rankine-Hugoniot 
formula owing to conductive heat flux cross the shock discontinuity. This has a 
significant effect on the properties of the post-shock gas in region 2 and the ambient 
flow in region 5.

Since the cloud evolution depends heavily on whether or not thermal conduction is 
efficient, we will treat these two regimes separately. The heat advection rate from 
thermal conduction is very sensitive to the temperature of the hot phase. Galaxy 
formation theory suggests that, in the real Universe, galactic haloes separate into 
cold haloes with gas at the photoionisation equilibrium temperature of $\sim10^4\kelvin$ 
and hot haloes at the virial temperatures over $10^{6}\kelvin$\citep{keres05, keres09a, 
dekel09}. Thermal conduction is, therefore, expected to be only important in hot haloes.

Magnetic fields can suppress thermal conduction significantly even if they are 
not dynamically important \citep{li19}. However, the strength of magnetic fields 
in galactic haloes is poorly constrained and the effects of tangled magnetic fields 
on thermal conduction is uncertain. Therefore, we do not explicitly model 
magnetic fields in this work. Instead, we use a free parameter to control the 
overall efficiency of thermal conduction.

To implement this process as a sub-grid model into cosmological simulations, we 
will focus on explicitly calculating the rate of deceleration and the mass loss 
rate of the cloud. The deceleration is caused by the ram pressure $\pii$ in front 
of the cloud and depends on the cross section $\pi\rcloud^2$ and the mass of the 
cloud $\mcloud$. Therefore, the deceleration rate is largely determined by the properties 
of the bow shock and of the compressed cloud, though both change drastically when 
one includes thermal conduction. 


The mass loss is primarily caused by hydrodynamic instabilities or thermal conduction 
or a combination of these two, but their calculation is more complicated. When thermal 
conduction is inefficient, the cloud loses its mass primarily from KHI and the 
expansion in the downstream direction after the cloud shock. The lifetime of the 
cloud is characterised by the Kelvin-Helmholtz time-scale $\tkh$, which will be 
described in \sect{sec:khi} based on the numerical results from \citetalias{sb15}. 
In the rest of the paper, we will mostly focus on the regimes where thermal conduction 
and evaporation are important. We develop a model for this regime based on the results 
from \citetalias{bs16}.

\begin{table*}
	\centering
	\caption{Numerical simulations from \citetalias{bs16} and their parameters.}
	\label{tab:simulations}
        \begin{minipage}{160mm}
	\begin{tabular}{lcccccccc} 
		\hline
		Name & $\mcloud/M_\odot$\footnote{The initial mass of the cloud.} & $\vrel [\mathrm{km s^{-1}}]$\footnote{The relative velocity of the cloud.} & $\rcloud [\mathrm{pc}]$\footnote{The initial radius of the cloud.} 
& $\ncloud [\mathrm{cm}^{-3}]$\footnote{The initial hydrogen number density of the cloud.} & $n_\mathrm{a} [\mathrm{cm}^{-3}]$\footnote{The hydrogen number density of the ambient medium} & $T_\mathrm{a} [\mathrm{K}]$\footnote{The temperature of the ambient medium} & $\Ncloud [\mathrm{cm}^{-2}]$\footnote{The initial column density of the cloud.} & $\tcc [\mathrm{Myr}]$\footnote{The cloud-crushing time-scale.} \\
		\hline
		$\chi300v1000$ & $6.7\times10^4$ & $1000$ & $100$ & $1.0$ & $3.3\times10^{-3}$ & $3\times10^6$ & $1.5\times10^{20}$ & $1.69$ \\
		$\chi300v1700$ & $6.7\times10^4$ & $1700$ & $100$ & $1.0$ & $3.3\times10^{-3}$ & $3\times10^6$ & $1.5\times10^{20}$ & $0.996$ \\
		$\chi1000v1700$ & $6.7\times10^4$ & $1700$ & $100$ & $1.0$ & $1.0\times10^{-3}$ & $10\times10^6$ & $1.5\times10^{20}$ & $1.82$ \\
		$\chi3000v3000$ & $6.7\times10^4$ & $3000$ & $100$ & $1.0$ & $1.0\times10^{-3}$ & $3\times10^6$ & $1.5\times10^{20}$ & $1.79$ \\
		$\chi300v3000$ & $6.7\times10^4$ & $3000$ & $100$ & $1.0$ & $0.33\times10^{-3}$ & $3\times10^6$ & $1.5\times10^{20}$ & $0.565$ \\
		$\chi300v3000b$ & $6.7\times10^4$ & $3000$ & $46.4$ & $10.0$ & $3.3\times10^{-3}$ & $3\times10^6$ & $1.5\times10^{21}$ & $0.262$ \\                                
		$\chi1000v480$ & $6.7\times10^4$ & $480$ & $100$ & $1.0$ & $1.0\times10^{-3}$ & $10\times10^6$ & $1.5\times10^{20}$ & $6.45$ \\                
		$\chi3000v860$ & $6.7\times10^4$ & $860$ & $100$ & $1.0$ & $0.33\times10^{-3}$ & $30\times10^6$ & $1.5\times10^{20}$ & $6.23$ \\                
		\hline
	\end{tabular}
\end{minipage}
\end{table*}

Both \citetalias{sb15} and \citetalias{bs16} study cloud evolution using a set of 
cloud-crushing simulations with varying flow parameters. \citetalias{bs16} includes 
isotropic thermal conduction at the Spitzer rate $\fS = 1$. Some of these simulations 
and their parameters are listed in \tab{tab:simulations}. The simulations are named 
after the initial density contrast $\chi$ and the relative velocity $\vrel$. These simulations 
explore a variety of physical conditions that are typical of interactions between 
winds and the hot CGM, with the ambient temperatures ranging from $3\times10^6\kelvin$ 
to $3\times10^7\kelvin$ and the initial Mach number ranging from 1.0 to 11.4. We 
also ran two additional simulations, $\chi300v1700c5$ and $\chi300v1700c20$ to explore 
the effects of reduced thermal conduction. They have the same initial conditions 
as $\chi300v1700$, but have only 1/5 and 1/20 of the original strength of thermal 
conduction, respectively.

\begin{figure*}
  \includegraphics[width=1.90\columnwidth]{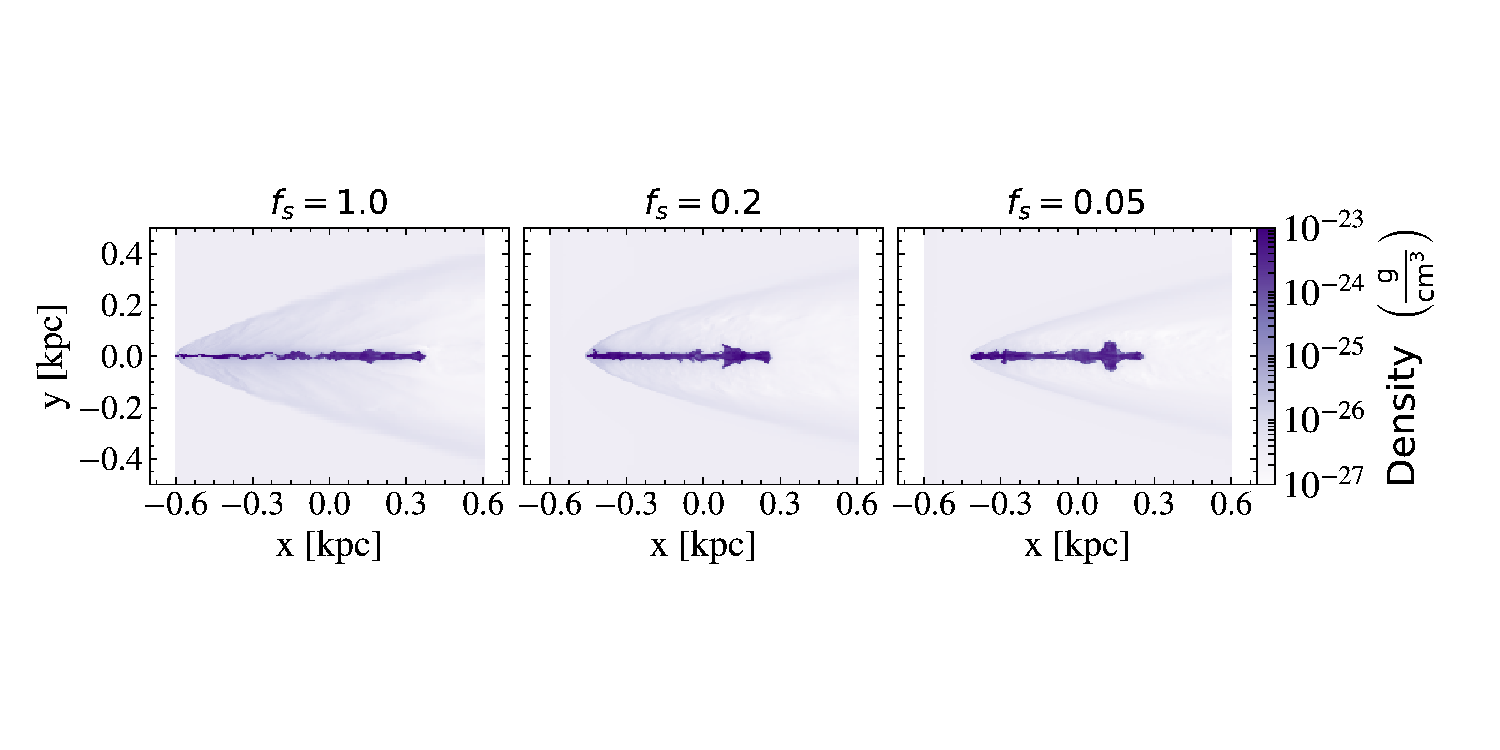}
  \centering
  \caption{From \textit{left} to \textit{right}: Density slices from the $\chi300v1700$, 
$\chi300v1700c5$ and $\chi300v1700c20$ simulations at $\sim 8.5$ Myrs. These simulations 
have different $\fS$ values as indicated in the figure. At this time, the cloud 
in the $\chi300v1700$ simulation has evaporated nearly 50\% of its mass, considerably 
more than in the other two simulations. The cloud in the $\chi300v1700c20$ simulation 
is still able to resist hydrodynamic instabilities even though thermal conduction 
is only 1/20 of the Spitzer value there.}
  \label{fig:density_lcond}
\end{figure*}



\section{Physical Processes}
\label{sec:physics}
In this section, we review the physical processes that are critical to the evolution 
of the cloud in the cloud-crushing problem and describe how to calculate relevant 
properties of the cloud and the ambient medium during its evolution. In \sect{sec:conduction_overview} 
we review some general formulae about thermal conduction. In \sect{sec:bow_shock} 
we find solutions to the bow shock and the cloud-crushing shock, with or without 
thermal conduction. In \sect{sec:expansion} we describe the morphology of the cloud 
after the cloud shock, the expansion of the cloud, and the internal structure of 
the cloud during the expansion. In \sect{sec:khi} we discuss how clouds lose mass 
owing to the Kelvin-Helmholtz instability and how to determine whether or not thermal 
conduction suppresses the KHI. In \sect{sec:conduction_zone} we propose an approximate 
model of estimating the mass loss rate from a cloud owing to conduction-driven 
evaporation. In \sect{sec:cooling}, we discuss the effects of radiative cooling 
on our analytic model.

\subsection{Classical and Saturated Conduction}
\label{sec:conduction_overview}
Throughout this paper we do not consider the effect of magnetic fields and assume 
isotropic thermal conduction. Thermal conduction relies on electrons in the hot 
plasma exchanging kinetic energy with the electrons in the cold gas. In the hot 
plasma, the mean free path of an electron is $\mfp \sim 1.65\times10^4\ [\cm]\ T_e^2/n_e$, 
where $T_e$ and $n_e$ are the electron temperature and electron number density in 
the plasma. In this paper, we will assume that electrons and ions are always in 
thermal equilibrium and have the same temperature $\Thot = T_e = T_i$ in the plasma, 
where $T_i$ is the ion temperature. In the classical limit where the mean free path, 
$\mfp$, is much smaller than the scale of the temperature gradient $L_\mathrm{T}$, 
thermal conduction leads to a heat flux
\begin{equation}
\label{eqn:q_class}
\qclass = -\kappahot\nabla T
,\end{equation}
where $\kappahot$ is a function of the temperature and density in the hot medium:

\begin{equation} 
\kappahot = 6.1\times10^{-7}\Thot^{5/2}\left( \frac{30}{\ln\Lambda_\kappa} \right)
\end{equation}
and $\ln\Lambda_\kappa \equiv 29.7 + \ln n_e^{-1/2}(T_e/10^6\kelvin)$ is the Coulomb 
logarithm that depends very weakly on $n_e$ and $T_e$. In this paper, we will always 
set $\ln\Lambda_\kappa = 30$ for simplicity.

When $\mfp \gg L_\mathrm{T}$, the cross sections of electron-electron collisions 
become too large for conduction to work in the classic limit. It reduces the efficiency 
of heat transfer to a saturated value \citepalias{cm77}:
\begin{equation}
\label{eqn:q_sat}
\qsat = 0.4\fS\left(\frac{2\kB T_e}{\pi m_e}\right)^{1/2}n_e\kB T_e \sim 1.715\times10^{-11}n_eT_e^{3/2}
,\end{equation}
where $\kB$ is the Boltzmann constant, $\fS \le 1$ is a free parameter that determines 
the overall efficiency of thermal conduction. $\fS = 1$ indicates thermal conduction 
at the Spitzer rate.

Similar to \citetalias{cm77}, we define a saturation parameter $\sigma_0$ that distinguishes 
classical conduction and saturated conduction based on the flow parameters:
\begin{equation}
\label{eqn:sigma0}
\sigma_0 = \frac{2\kappahot\Thot}{25\rho_hc_{h}^3\rcloud} = 1.84\frac{\lambda_h}{L_T}
,\end{equation}
where $c_h$ is the isothermal sound speed of the hot medium, and 
$\lambda_h$ is the mean free path of the hot medium. Physically, $\sigma_0$ 
is the ratio between $\qclass$ and $\qsat$. We use the classical heat flux when 
$\sigma_0 < 1$ and the saturated heat flux otherwise.

Thermal conduction at an interface between hot and cold gas could lead to evaporation 
of cold gas into the hot gas, as the cold gas near the interface gains energy from 
electron collision. In the classical limit, \citetalias{cm77} derive the evaporation 
time-scale for a spherical cloud of uniform density in an initially uniform, infinite 
hot medium as
\begin{equation}
\tau_\mathrm{ev,class} = \frac{25\kB \mcloud}{16\pi \mu \mh \kappa \fS \rcloud}
,\end{equation}
where $\mu$ is the atomic weight and $\mh$ is the mass of the hydrogen atom. The 
above equation can be written numerically as:
\bq
\label{eqn:tevap_class}
\tau_\mathrm{ev,class} = 48.9\ [\mathrm{Myr}]\left[ \left(\frac{\ncloud}{1\,\cm^{-3}}\right) \left(\frac{\rcloud}{10\parsec}\right)^2 \left(\frac{\Thot}{10^6\kelvin}\right)^{-2.5} \right]
,\eq
where $\rcloud$ and $\mcloud$ are the radius and mass of the spherical cloud, and 
$\fS \le 1$ is a constant factor that determines the strength of thermal conduction 
relative to the Spitzer value (\eqn{eqn:q_class}).

In the saturated limit ($\sigma_0 > 1$), the evaporation time-scale becomes \citepalias{cm77}:
\bq
\label{eqn:tevap_sat}
\tau_\mathrm{ev,sat} = 10.3\ [\mathrm{Myr}] 
\left(\frac{\chi}{10^3}\right)
\left(\frac{\rcloud}{10\parsec}\right)
\left(\frac{\Thot}{10^6\kelvin}\right)^{-1/2}
\sigma_0^{-3/8}
,\eq
which is obtained from their equation 64 with the parameter $\phi_\mathrm{s}$ in 
the equation set to 1.0.

Note, however, that the above treatment of evaporation is only valid when the mean 
free path of hot electrons inside the cloud is much smaller than the cloud radius. 
Otherwise, hot electrons will be able to free stream through the cloud while at 
the same time heating the entire cloud through coulomb heating \citep{balbus82}. 
When the coulomb heating rate exceeds the radiative cooling rate, the cloud will 
puff up quickly and disintegrate shortly thereafter \citep{li19}. This only occurs 
for very small clouds in a very hot medium and puts a lower limit on 
the initial cloud size, which is a main parameter of our model. \citetalias{bs16} 
show that this quick disruption occurs when the initial column density of the cloud 
is smaller than $1.3\times10^{18}\cm^{-2}(T_1/10^7\kelvin)^2$ and demonstrate 
in a test simulation that a cloud with an initial size of $\rcloud = 1\parsec$ in a 
surrounding medium with $\chi=3000, T_1=10^7\kelvin$, and $\vrel=3000$\kms indeed 
evaporates within $1\tcc$. However, this lower limit is much below the physical 
conditions probed in \citetalias{bs16} in which we are interested in this paper.

The \citetalias{cm77} solution for conductive evaporation also assumes that radiative 
cooling is negligible. We will further discuss the effects of cooling on thermal conduction 
in \sect{sec:cooling}.

\subsection{The Bow Shock and the Cloud Shock}
\label{sec:bow_shock}



\subsubsection{Without Thermal Conduction}

In the non-conductive limit, we approximate the bow shock as adiabatic so that the 
physical conditions at the two sides of the shock (boundary I) are related by the 
Rankine-Hugoniot jump conditions (appendix~\ref{sec:shock_jump_conditions}). In 
the post-shock gas (region 2), the flow is subsonic and is governed by the Bernoulli 
equations that relate post-shock quantities at boundary I to the fluid quantities 
at the stagnation point II. 
The pressure at the stagnation point and that in the shocked cloud (region 4) are 
the same ($\pii = P_4$).

The thermal pressure of the pre-shock medium $P_1$ and the pressure at the stagnation 
point is, therefore, related by \citep{mc75}\footnote{The formula only applies to the pressure in front of the cloud. The pressure behind the oblique shock is smaller than this value. In this paper, we only consider the pressure resulting from the front shock.}:

\begin{equation}
\label{eqn:P_II}
\frac{\pii}{P_1} = \frac{\pii}{P_{I}}\frac{P_{I}}{P_1} = \left( \frac{\gamma + 1}{2}\right)^{\frac{\gamma + 1}{\gamma - 1}}\left( \gamma - \frac{\gamma - 1}{2\mathcal{M}_1^2}\right)^{-\frac{1}{\gamma - 1}}\mathcal{M}_1^2
,\end{equation}
where $\mathcal{M}_1$ is the Mach number of the pre-shock gas in the velocity frame of the cloud:
\bq
\mathcal{M}_1 \equiv \frac{v_1}{\csone}
\eq
and $\csone$ is the isothermal sound speed\footnote{The isothermal sound speed is 
defined as $c_\mathrm{iso}^2 = \gamma(P/\rho) = \gamma \kB T/(\mu \mh)$.} in the 
unshocked gas.

\eqn{eqn:P_II} only applies in the supersonic case $\mathcal{M}_1 > 1$. Here $\pii$ is also the ram pressure that is responsible for the deceleration of the cloud:


\begin{equation}
\label{eqn:P_ram}
\pram \sim \fram(\mathcal{M}_\mathrm{rel})\rho_\mathrm{a}\vrel^2
,\end{equation}
where $\mathcal{M}_\mathrm{rel}$ is the mach number of the ambient flow relative to the cloud, and the coefficient $\fram$ can be derived from \eqn{eqn:P_II} with $\pram=\pii$. 
It is of order unity and has a minimum value 0.5 when $\mathcal{M}_\mathrm{rel} = 1$. 
For simplicity, we choose $\fram = 0.5$ in this paper. Our results are not sensitive to this choice of $\fram$.

The cloud shock propagates at a speed $\vshock$ that can be solved using the jump 
conditions at the cloud shock front. Assuming the cloud shock is isothermal, we 
approximate the shock speed according to the jump condition:

\begin{equation}
\frac{\pii}{P_3} = \left( \frac{\vshock}{c_\mathrm{s,c}} \right)^2
,\end{equation}
where $c_\mathrm{s,c}$ is the isothermal sound speed of the cloud. The shock speed 
is related to the cloud crushing time by $\vshock \sim \rcloud(t=0)/\tcc$.

Therefore, by assuming an adiabatic bow shock and an isothermal cloud shock, we 
are able to solve for the post-shock properties of the cloud. The cloud is compressed 
within a few $\tcc$ and accelerated to the shock speed $\vshock$. The density inside 
the cloud is enhanced by a factor of $\rho_4/\rho_3 \sim \chi^{-1}(v_1/c_\mathrm{s,c})^2$, 
making it over-pressured relative to its surroundings.

\subsubsection{With Conduction}
\label{sec:shock_conduction}
Including thermal conduction could significantly affect the bow shock as well as the 
cloud-crushing shock. Either in the regime of classical conduction, where $q \sim 
T^{2.5}$, or in the regime of saturated conduction, where $q \sim n_eT_e^{1.5}$, 
the heat flux $q$, the evaporation rate $\dot{m}$, and the vapour pressure $\pevap$ 
are all very sensitive to the post-shock properties of the flow. Furthermore, the 
post-shock flow is no longer a constant flow determined by the Rankine-Hugoniot 
jump condition, but rather displays a time-dependent profile behind the main shock 
front. The picture of a radiative shock with electron thermal conduction has been 
extensively studied in the literature \citep{lacey88, borkowski89}. While these 
works focus on plane-parallel shocks driven by a supersonic flow in a single continuous 
medium, the cloud-crushing problem requires a self-consistent solution in a two-phase 
medium, i.e., hot ambient gas and a cool cloud.

Despite these complications, we extended the \citet{borkowski89} prescription for 
a conductive shock by including a non-negligible initial temperature. Inheriting 
their notation, we may solve for the modified Rankine-Hugoniot jump conditions (see 
appendix~\ref{sec:shock_jump_conditions} for the derivation). The density and temperature 
ratio across the conductive shock front becomes:
\begin{equation}
\label{eqn:jc_density_ratio}
\xs \equiv \frac{\rho_1}{\rho_2} = \frac{5(1+\betas) - \sqrt{9+16\qs+5\betas(5\betas-6)}}{8}
\end{equation}
and
\begin{equation}
  \label{eqn:jc_temperature_ratio}
  \frac{T_1}{T_2} = \frac{\betas}{(1+\betas-\xs)\xs}
.\end{equation}

In the above equations, $\qs$ is a parameter that is explained in detail below, 
and $\betas$ is defined as:
\begin{equation}
\label{eqn:jc_beta}
\betas \equiv \frac{1}{\gamma \mach_1^2}
.\end{equation}

In the extreme case where $\mach_1 \gg 1$, the equations reduce to equation 16 and 
equation 17 in \citet{borkowski89}.

Equations \ref{eqn:jc_density_ratio}, \ref{eqn:jc_temperature_ratio} and \ref{eqn:jc_beta} 
introduce a parameter $\qs$, which we define as the ratio between the conductive heat flux 
and the kinetic energy flux of the incoming flow across the shock:
\begin{equation}
\label{eqn:qs}
\qs \equiv \frac{q_\mathrm{s}}{\frac{1}{2}\rho_1v_1^3}
.\end{equation}

$\qs$ measures how much of the thermal energy generated in the shock is advected back into the pre-shock gas. $\qs = 0$ corresponds to an adiabatic shock and $\qs = 1$ corresponds to an isothermal shock. For any given pair of $\qs$ and $\mach$, the density and temperature ratios between the post-shock gas and the pre-shock gas are uniquely determined by Equations \ref{eqn:jc_density_ratio} and \ref{eqn:jc_temperature_ratio}. 

\begin{figure}
  \includegraphics[width=0.95\columnwidth]{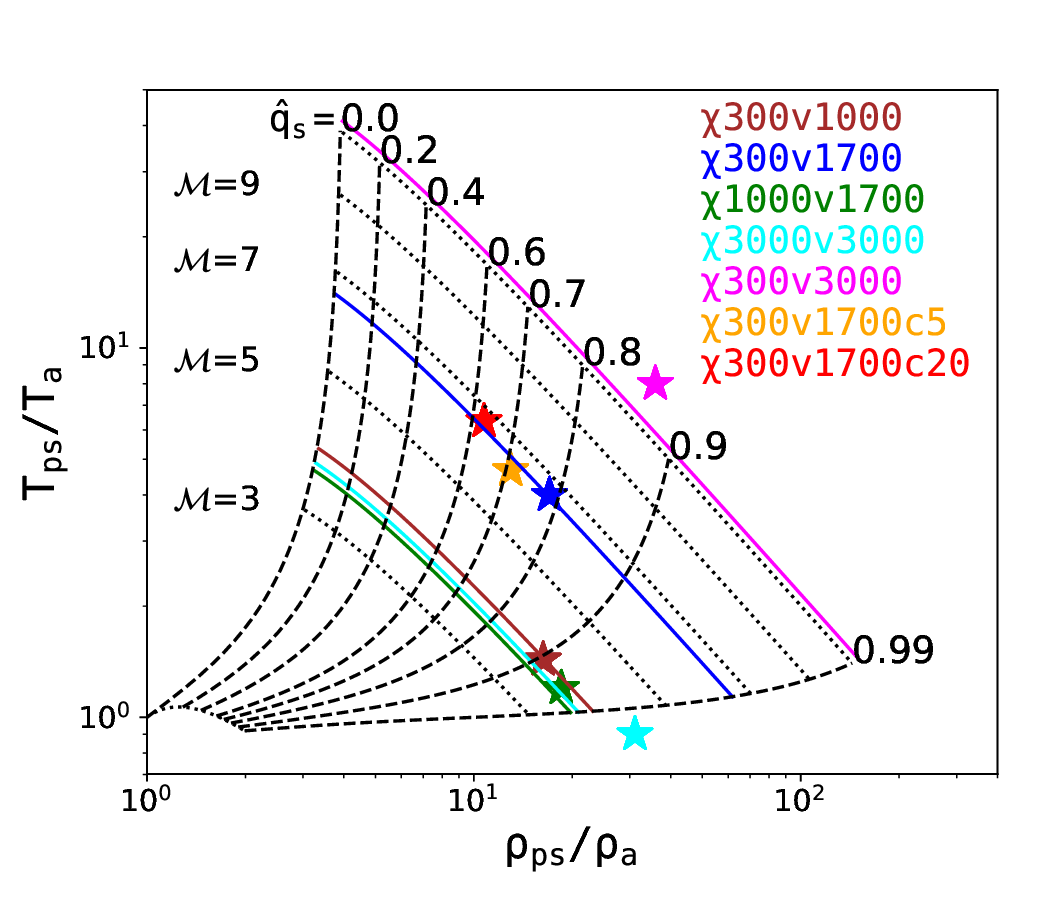}
  \centering
  \caption{The ratio between the post-shock density and temperature, i.e., $\rho_\mathrm{ps}$ 
and $T_\mathrm{ps}$, and the pre-shock properties, i.e., $\rho_\mathrm{a}$ and $T_\mathrm{a}$, 
calculated for different sets of $(\qs, \mach)$ pairs using the conductive jump 
conditions (\eqn{eqn:jc_density_ratio} and \eqn{eqn:jc_temperature_ratio}). The 
$\mach$ number is constant on each dotted line and $\qs$ is constant on each dashed 
line. The stars correspond to measurements from \citetalias{bs16} simulations, with 
the colours indicating particular numerical models as indicated in the figure. We 
also show solid lines that correspond to the $\mach$ number of each simulation. Since the $\chi300v1700$, $\chi300v1700c5$ and $\chi300v1700c20$ simulations use the same $\mach = 6.46$, we use a single blue line to indicate all three simulations.}
  \label{fig:qs}
\end{figure}

However, the exact value of $\qs$ varies among simulations and is hard to determine 
from first principles. We measure these ratios from the \citetalias{bs16} simulations 
at $\tninty$, when the cloud reaches 90\% of its original mass, and compare them to the analytic solutions in \fig{fig:qs}.
We find that the measured ratios lie close to the lines that are defined by their 
corresponding Mach number. However, $\qs$ varies among simulations that have a similar 
Mach number. In general, when thermal conduction is strong, as in the $\chi3000v3000$ 
simulation where the ambient temperature is very high, $\qs$ is closer to unity, 
corresponding to a nearly isothermal shock. This is expected because, as the width 
of the bow shock develops over time, the temperature gradient after the shock gradually 
declines and the shock profile approaches an isothermal one. On the other hand, 
when thermal conduction decreases, as from the full Spitzer value $\fS$ in $\chi300v1700$ 
to $0.05\fS$ in the $\chi300v1700c20$ simulation, $\qs$ decreases.

In our model, we choose a constant $\qs = 0.90$ whenever thermal conduction 
is non-negligible for simplicity. 

In light of the conductive simulations from \citetalias{bs16}, we further assume 
that the cloud will always have a cylindrical geometry after being compressed. 
We set the dimensions of the cloud as $(\pi \rcloud^2) \times \lcloud$, where $\rcloud$ 
is the cross section of the cloud perpendicular to the flow and $\lcloud$ is the 
length of the cloud parallel to the flow. We set $\lcloud = 2\rcloud$ immediately 
after the shock and can solve for $\rcloud$ once the density of the compressed cloud 
$\rho_4$ is known. We also choose a coordinate system such that the $x-$axis is 
the central axis of the cylinder with the origin $x=0$ at the head of the cloud 
(\fig{fig:conduction_zone}). 

\subsection{Expansion}
\label{sec:expansion}
After maximum compression from the cloud shock, the cloud expands rapidly in the 
downstream direction into a nearly vacuum cavity that is enclosed by the surface 
extended from the contact discontinuity. In simulations without thermal conduction \citepalias{sb15}, 
the expansion flow is often strongly perturbed by the ambient flow and quickly mixes 
into the ambient medium. In addition, a Rayleigh-Taylor instability at the front 
of the cloud often breaks up the cloud into smaller clumps, making the mixing process 
even more efficient. Therefore, in the non-thermal conduction regime, the clouds 
often do not have a well defined morphology.

In simulations with thermal conduction \citepalias{bs16}, the clouds often display 
a coherent, cylindrical morphology (e.g., see \fig{fig:density_lcond}). When thermal 
conduction is strong enough, it helps suppress hydrodynamic instabilities and confine 
the cloud with vapour pressure. Simulations also show a strong velocity gradient 
within the cloud throughout its expansion. In the velocity frame of the contact 
point at II, the expansion velocity increases linearly with the distance to the 
contact point and reaches a maximum at the tail of the cloud, where the cloud gas 
almost freely flows into the cavity with a speed comparable to the shock velocity 
$\vshock$. However, when thermal conduction become less efficient, the cloud morphology 
becomes less stable and eventually the cloud breaks up faster.

Therefore, we will only approximate the cloud as a cylinder when thermal conduction 
is sufficiently strong (see \sect{sec:khi} for more details on determining whether 
or not this is true). In our model with thermal conduction, it is important to know 
how the length of the cloud evolves with time, because the total evaporation rate 
from the cloud depends on the total surface area, i.e., $2\pi\rcloud\lcloud$, 
of the cloud at any time. 

Immediately after the time of maximum compression, the velocity structure inside 
the simulated clouds resembles a similarity flow \citep{landau59}, with the velocity 
at any point $x$, $v(x)$, increasing linearly with $x$. The flow in the cloud is 
a centred rarefaction wave until the wave propagates back to the location of the 
bow shock. The simulations show that the tail of the cloud often expands at a nearly 
constant velocity, $\vexp$, so that the cloud length grows as $d\lcloud/dt = \vexp$. 
If the expansion is adiabatic, the cloud should expand at a terminal velocity $v_\mathrm{max,ad} 
= 2\cscloud/(\gamma-1) \sim 45$\kms as expected from an adiabatic similarity flow. 
However, the $\vexp$ measured from the simulations 
is often much larger than this value. Here, we assume the expansion is isothermal. 
For an isothermal similarity flow, the density $\rhocloud(x)$ and pressure $\pcloud(x)$ at any position 
$x$ inside the expanding cloud are functions of the flow velocity $v(x)$ only:
\begin{equation}
\label{eqn:crw}
\frac{\rhocloud(x)}{\rhocloud(0)} = \frac{\pcloud(x)}{\pcloud(0)} = \exp \left( -\frac{v(x)}{\cscloud}\right)
,\end{equation}
where $\rhocloud(0)$ and $\pcloud(0)$ are the density and pressure at the head of 
the cloud. Since the velocity $v(x)$ in an isothermally expanding cloud increases with 
$x$ and does not have an upper limit, we need to arbitrarily choose a maximum velocity 
as $\vexp$, which corresponds to the velocity at the tail of the cloud. 
\eqn{eqn:crw} indicates that the cloud segment with a larger $v$ has a lower density 
and evaporates faster. Therefore, the further away from the head, the faster the cloud 
evaporates. In our model, we choose $\vexp$ as the velocity at which the cloud still 
has not fully evaporated. At any time $t$, the fraction of the cloud where 
$v > \vexp$, i.e., $\rho(v) < \rho(\vexp)$, has evaporated earlier. The $\vexp$,
therefore, decreases with time as
\bq
\label{eqn:v_exp_evap}
\vexp\mathrm{(ev)}(t) = -\cscloud\ln \left( 4.5\times10^{-15}\frac{T_1^{5/2}t}{\rhocloud(0)\rcloud^2} \right)
,\eq
where we use \eqn{eqn:mlra_class} to find the evaporation rate per unit area $\mlra$ 
for classical conduction and use the temperature for the unperturbed ambient flow. The material that has velocities that exceed the expansion velocity are assumed to have evaporated.

On the other hand, we can choose $\vexp$ as the velocity at which the cloud pressure 
equals the pressure of the unperturbed ambient, i.e., $\pcloud(\vexp) = P_1$. Since 
the pressure at the head of the cloud, $\pcloud(0)$ equals the ram pressure $\pii$, 
the expansion velocity is:
\begin{equation}
\label{eqn:v_exp_P}
\vexp\mathrm{(P)} = -\cscloud\ln\left(\frac{P_1}{\pii}\right)
\end{equation}

In practice, we choose the minimum value of these two velocities as the expansion 
velocity in our model:
\bq
\label{eqn:v_exp}
\vexp = \mathrm{min}\{\vexp\mathrm{(ev)}, \vexp\mathrm{(P)}\}.
\eq


\subsection{The Kelvin-Helmholtz Instability}
\label{sec:khi}
The growth rate of perturbations at the interface of a shearing flow is characterised 
by the Kelvin-Helmholtz time-scale $\tkh$. A classical analysis in the subsonic, 
incompressible limit shows that $\tkh \propto \tcc$ for linear growth \citep{chandrasekhar61, 
mandelker16}. In a supersonic flow, the KHI is damped, but the exact behaviour is 
poorly understood. Moreover, it is not straightforward to apply the classic $\tkh$ 
to the cloud-crushing problem, where the geometry and long term evolution are distinct 
from those assumed in the classical analysis of the KHI. Radiative cooling also 
has a strong effect on the growth of the KHI (see \sect{sec:cooling} for details). 
Using their non-conductive simulations, \citetalias{sb15} obtain an empirical result 
for how fast the cloud loses mass in various situations. They find that the times 
at which the cloud has a certain fraction, e.g., 90\%, 75\%, 50\%, 25\%, of its 
original mass are proportional to $\tcc\sqrt{1+\mach_\mathrm{h}}$ (their equation 
22). The additional $(1+\mach_\mathrm{h})^{1/2}$ factor suggests that the clouds 
survive much longer in highly supersonic flows than that predicted from a classic 
analysis. Therefore, we adopt the following formula for clouds in regimes where 
thermal conduction is negligible:
\bq
\label{eqn:tkh}
\tkh = \fkh\sqrt{1+\mach_1}
,\eq
where $\fkh$ is a free parameter of order unity that controls how fast clouds lose 
mass via KHI, and $\mach_1 = \vrel / c_1$ is the Mach number of the flow relative 
to the cloud. We calculate the mass loss rate of the cloud as:
\bq
\label{eqn:mlr_kh}
\dot{M}_\mathrm{c;KH} = \frac{\mcloud}{\tkh}
.\eq

Whether or not KHI can grow depends also on the strength of thermal conduction. 
In the extreme case where evaporation dominates over the ambient flow, it simply 
eliminates any velocity shear. With less strong thermal conduction, linear perturbations 
on the cloud surface can still be stabilised if the kinetic energy diffuses quickly 
enough before it can generate a significant amount of local vorticity.

A full treatment of this problem requires solving the linearly perturbed equations 
that include a conductive flux term in the energy equation, which is very challenging 
even in ideal situations. Here we derive an approximate criterion based on whether 
or not the diffusion time-scale owing to thermal conduction, $\tdiff$, is shorter 
than the mixing time-scale, $\tmix$, owing to KHI.

Consider a hot phase with density $\rhohot$ and temperature $\Thot$ flowing at a 
relative velocity of $\vrel$ to a cold phase with density $\rhocloud=\chi\rhohot$ 
and temperature $\Tcloud$. A perturbation on the scale of $\lambda$ in the cloud 
will mix into the ambient flow over a finite width $\lambda$ within $\tmix$ without 
thermal conduction. We can obtain the mixing time-scale using the dispersion relation 
for the growth of linear perturbations \citep{mandelker16}:
\bq
\tmix \sim \chi^{1/2}\lambda/\vrel.
\eq
To calculate the diffusion time-scale, we consider how long it takes the conductive 
heat flux to fully mix the kinetic and thermal energy between a density perturbation 
with its surroundings on any scale $\lambda$: 
\bq
\tdiff = \frac{1}{1+\chi}\frac{\rhocloud\lambda[\vrel^2/2 + 3\chot^2/2]}{q}
,\eq
where $\chot$ is the sound speed of the hot gas, the factor $1/(1+\chi)$ is the 
volume filling factor for the cold phase, and $q$ is the conductive heat flux. Here 
we assume classical conduction and approximate it as $q = \fS\kappahot(\Thot/\lambda)$. 
There exists a critical scale $\lkh$ where $\tdiff = \tmix$. Perturbations are able 
to grow only on scales smaller than $\lkh$:
\bq
\label{eqn:lambda_kh}
\lkh = \frac{1+\chi}{\chi^{1/2}}\frac{1}{\machhot(\machhot^2+3)}
\frac{\fS\kappahot}{\nhot\Thot^{1/2}}
\left(\frac{4\mu\mh}{\gamma^3\kB^3}\right)^{1/2}
,\eq
where $\nhot$ and $\machhot$ are the hydrogen number density and the Mach number 
of the hot phase, respectively. In the cloud-crushing problem, the KHI is able to 
grow only when $\rcloud > \lkh$. Using $\chi \gg 1$, and $\machhot \sim 1$ for the 
post-shock gas, we can write $\lkh$ numerically as:
\bq
\lkh = 5.7\ [\mathrm{kpc}]\
\fS
\left( \frac{\chi}{10^2}\right)^{1/2}
\left( \frac{\Thot}{10^7\kelvin}\right)^2
\left( \frac{\nhot}{10^{-2}\ \mathrm{cm^{-3}}} \right)^{-1}.
\eq

 \citet{mandelker16} find a similar dependence of $\lkh$ on fluid properties, i.e., $\lkh \propto \Thot^2\nhot^{-1}\machhot^{-1}$. For most of the simulations from \citetalias{bs16} with full Spitzer rate conduction, i.e., $\fS = 1$, the critical length $\lkh$ is much larger than the cloud radius $\rcloud$, so that the Kelvin-Helmholtz instabilities are always suppressed. 

However, when one reduces $\fS$, KHI will eventually be able to grow. In the three 
$\chi300v1700$ simulations, using properties of the ambient flow of $\nii\sim0.1\ 
\mathrm{\cm^{-3}}$, $\Tii\sim7\times10^6\kelvin$, and $\chi\sim10^3$, we find that 
the critical scales for the $\chi300v1700$, $\chi300v1700c5$, and the $\chi300v1700c20$ 
simulations are 890 pc, 178 pc, and 45 pc, respectively. Only in the $\chi300v1700c20$ 
simulation is the critical scale comparable to the cloud radius $\rcloud\sim20\parsec$, 
and this is the only simulation that indeed shows some growth of the KHI at later 
times that ultimately breaks up the cloud. In \sect{sec:tests} we will show that 
KHI indeed causes the cloud to lose mass in addition to evaporation.


\subsection{The Conduction Zone}
\label{sec:conduction_zone}

\begin{figure}
  \includegraphics[width=0.95\columnwidth]{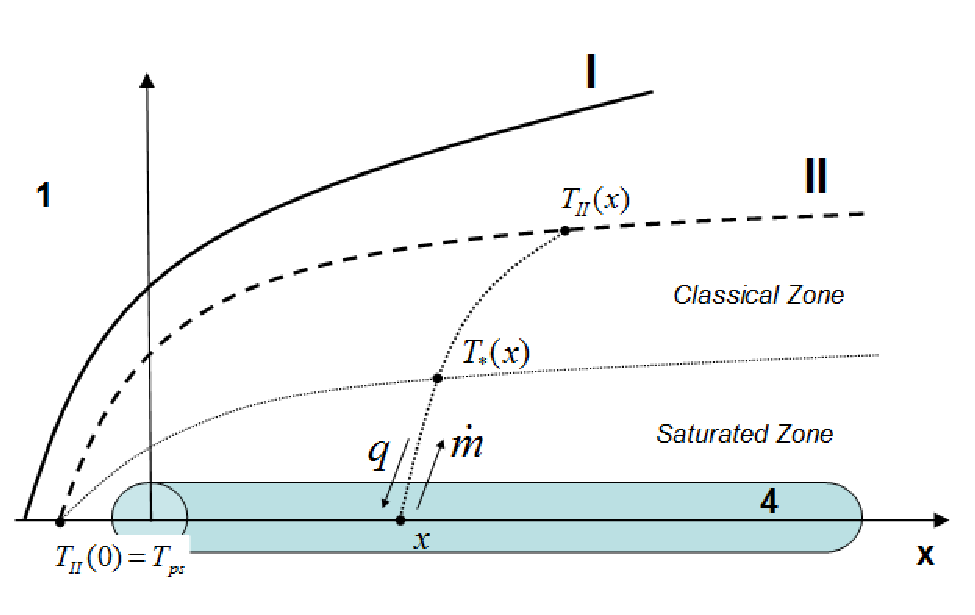}
  \centering
  \caption{An illustration of our method of calculating the conductive evaporation 
rate from the cloud at any given time. We assume cylindrical symmetry. Roman and 
Arabic numerals have the same meanings as in \fig{fig:cartoon}. We assume that thermal 
conduction between the cloud and the ambient flow causes heat flux and evaporation 
within a conduction zone bounded by the surface of the cloud and an arbitrarily 
chosen streamline (noted as Roman numeral II) in the ambient flow. Depending on 
the strength of the conduction, the region may be divided into a classic zone and 
a saturated zone separated by a transition surface, where the quantities are subscribed 
with a star symbol (*).}
  \label{fig:conduction_zone}
\end{figure}

When thermal conduction is strong, cold gas evaporates from the cloud surface and 
mixes into the ambient flow moving downstream. To solve for the mass loss rate from 
the cloud, we assume that the flow is axisymmetric and that there exists a continuous 
conduction zone (see \fig{fig:conduction_zone}) extending from the cloud surface 
III to an arbitrary surface II in the ambient flow. Inside the conduction zone, 
the gas that evaporated from any coordinate $x$ in the cloud is heated from the 
cloud temperature $\Tcloud$ to a corresponding ambient temperature at the surface 
II, i.e., $\Tii(x)$. The temperature varies along the streamlines, dropping from 
the maximum value at the shock front $T_{ps}$ (or $T_2$ as in \fig{fig:cartoon}), 
to the unperturbed ambient temperature $T_1$ far behind the shock. We now focus 
on streamlines (dotted lines) along which the evaporated material flows. Each of 
these paths relates fluid properties at one point on the cloud to those at another 
point on the surface II. We approximate these streamlines of evaporated material 
as radial to the clouds so that we can analytically integrate over the radial coordinate 
$r$ from the cloud surface $r=\rcloud$ to the ambient $r=\rii$. The problem is to 
find an approximate expression for $\Tii(x)$, parameterized by the cloud coordinate 
$x$, and to find the mass loss rate per unit area $dA = 2\pi\rcloud dx$ at any $x$ 
of the cloud, defined as:
\begin{equation}
  \label{eqn:mlra}
  \mlra = 2\pi r\rho v = \mathit{const.}
\end{equation}

\begin{figure}
  \includegraphics[width=0.95\columnwidth]{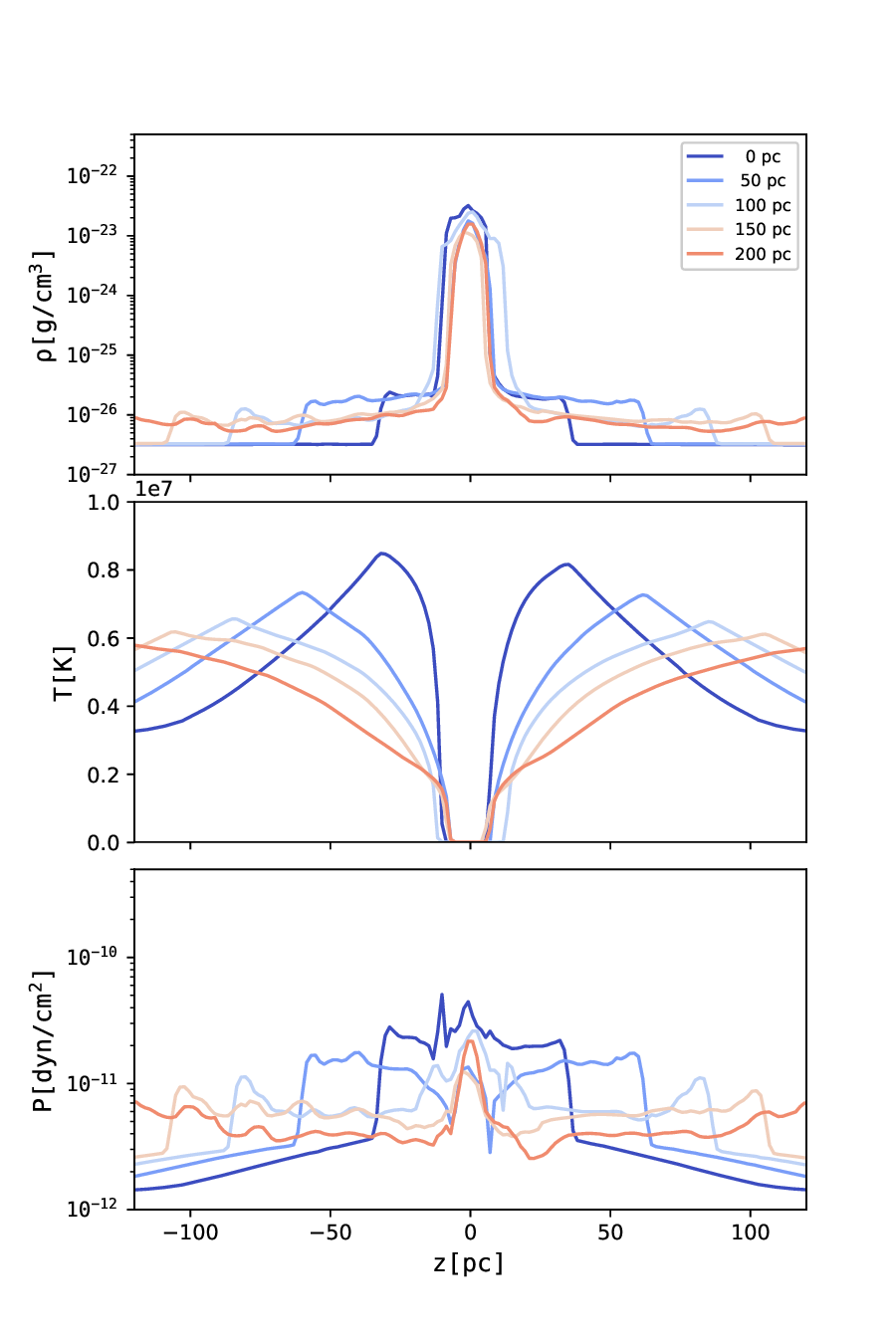}
  \centering
  \caption{The radial profiles of density (\textit{upper panel}), temperature (\textit{middle 
panel}) and pressure (\textit{lower panel}) at different cross sections of a simulated 
cloud. The data is from the $\chi300v1700$ simulation at $\tseventyfive$. The legend 
shows how far the cross section is from the head. Note that the temperature axis 
in the middle panel is shown in linear scale, so it's clear that the temperature 
gradient sharpens towards the cloud, making thermal conduction more likely to saturate.}
  \label{fig:cond_zone_x300v1700}
\end{figure}

We show a typical conduction zone from the simulations in \fig{fig:cond_zone_x300v1700}. 
The profiles show three distinct regions, separated by two sharp density discontinuities. 
From inside out, the three regions correspond to the cloud, the post-shock ambient 
flow (conduction zone), and the flow outside the bow shock. The conduction zone 
broadens with the distance from the head (from blue to red), consistent with the 
morphology illustrated in \fig{fig:cartoon}. The lateral dimension of the cloud, 
i.e. $\rcloud$, varies little along the cloud. 

In the conduction zone, the temperature gradient sharpens towards the cloud surface, 
where the conductive flux will likely start to saturate. When thermal conduction 
is strong enough, there exists a critical point where the heat flux starts to saturate 
so that it divides the conduction zone into a classic zone and a saturated zone, 
which will be treated separately below. The flow properties at the critical point 
are noted as $\rho_*$, $T_*$, etc.

In both regions, the flow along any path is governed by the time-independent Euler 
equations in cylindrical coordinates:
\bq
\label{eqn:euler1}
\rho\frac{dv}{dr} + v\frac{du}{dr} + \frac{\rho v}{r} = 0,
\eq
\bq
\label{eqn:euler2}
\rho v \frac{dv}{dr} = -\frac{dp}{dr},
\eq
and
\bq
\label{eqn:euler3}
\frac{3}{2}\gamma\mlra(1+\frac{1}{5}\mach^2)c^2 = 2\pi r q.
\eq

All flow quantities in the above equations are functions of $r$. The heat flux $q$ 
is determined by either \eqn{eqn:q_class} or \eqn{eqn:q_sat} in the classical and 
the saturated zone, respectively. In the classical zone, $\mach^2/5 \ll 1$ so that 
$\mlra \propto r\qclass/T$. Since $\mlra$ is constant along the streamline, $\qclass$ 
is proportional to $T/r$, which decreases with $r$. There might exist a critical 
point $r=r_*$, where $\qsat = \qstar = \qclass$. At $r < r_*$, $\sigma_0 > 1$, this 
corresponds to saturated conduction, while at $r > r_*$, $\sigma_0 < 1$, this corresponds 
to classical conduction. We further define $\sigma_*$ and $\sigmacloud$ as the value 
of $\sigma_0$ at $r=r_*$ and $r=\rcloud$, respectively. By definition, $\sigma_* 
= 1$. Therefore, the critical point exists if and only if $\sigmacloud > 1$.

In the classical zone, we could obtain $\mlra$ by integrating the energy equation 
(\eqn{eqn:euler3}) from $r=r_*$ to $r=\rii$ using \eqn{eqn:q_class} with the approximation 
that $1+\mach^2/5 \sim 1$:
\bq
\label{eqn:mlra_class}
\mlra\mathrm{(class)} = 6.1\times10^{-7}\fr^{-1}
\left(\frac{8\pi\mu\mh}{15\gamma\kB}\right)
(\Tii^{5/2}-T_*^{5/2})
,\eq
where $\fr \equiv \ln(\rii/\rcloud)$ is of order unity. We will use $\fr = 1$ in 
this paper.

In the saturated zone, integrating the energy equation \eqn{eqn:euler3} shows that 
the Mach number has a constant value $\machsat$ that is determined by
\bq
\label{eqn:mach_sat}
\machsat(1+\frac{1}{5}\machsat^2) = 2\fS
,\eq
where $\machsat \sim 1.4$ throughout the saturated zone for thermal conduction at 
the full Spitzer rate, i.e., $\fS = 1$, and becomes smaller with a reduced $\fS$.

With a constant $\machsat$, we can solve for the temperature profile in the saturated 
zone by integrating the continuity equation (\eqn{eqn:euler1}) and the equation 
of motion (\eqn{eqn:euler2}) in cylindrical coordinates:
\bq
\label{eqn:temperature_profile_sat}
\left(\frac{T}{\Tcloud}\right) = 
\left(\frac{r}{\rcloud}\right)^{2/(1+\machsat^2)}.
\eq

To provide a boundary condition at $r_*$ for the saturated zone, we further assume 
that the pressure gradient in the classical zone is negligible as indicated by the 
simulation (\fig{fig:cond_zone_x300v1700}), so that
\begin{equation}
\label{eqn:cond_zone_pressure}
n_*T_* = \nii\Tii
.\end{equation}

We can solve for the mass loss rate in the saturated zone with \eqn{eqn:euler3}, 
\eqn{eqn:mach_sat} and \eqn{eqn:cond_zone_pressure}:
\bq
\label{eqn:mlra_sat}
\begin{split}
  \mlra\mathrm{(sat)} = &\,1.715\times10^{-11}
  \left( \frac{4\pi}{3+\machsat^2} \right)
  \left( \frac{\mu\mh}{\gamma\kB} \right)\times\\
  &\,T_*^{-1/2}
  \left( \frac{\nii\Tii}{\rcloud} \right)
  \left( \frac{\Tcloud}{T_*} \right)^{(1+\machsat^2)/2}.
\end{split}
\eq

We can find the temperature at the critical point $T_*$ by iteratively solving the 
equation $\sigmastar = 1$, where the saturation parameter, $\sigmastar$, is by definition 
the ratio $\qclass/\qsat$ at $r_*$:
\bq
\label{eqn:cond_zone_saturation_parameter}
\begin{split}
  \sigmastar = &\,3.6\times10^4\fr^{-1}
  \left( \frac{6+2\machsat^2}{15} \right)\times\\
  &\,\left[ \frac{(\Tii^{5/2}-T_*^{5/2})T_*^{1/2}}{\nii\Tii\rcloud} \right]
  \left( \frac{\Tcloud}{T_*} \right)^{(1+\machsat^2)/2}.
\end{split}
\eq

A solution for $T_*$ is physical only if $T_* \ge T_c$. It is clear from \eqn{eqn:cond_zone_saturation_parameter} 
that $\sigmastar$ increases with $r$ so that $\sigmastar \ge \sigmacloud$. Therefore, 
the criterion that a saturated zone exists is $\sigmacloud \ge 1$. Using a fiducial 
set of parameters, $\fr = \fS = 1$, this criterion becomes:

\bq
\label{eqn:cond_zone_criteria}
\sigmacloud = 2.4\times10^4
\left[ \frac{(\Tii^{5/2}-\Tcloud^{5/2})\Tcloud^{1/2}}{\nii\Tii\rcloud} \right]
\ge 1.
\eq

We then obtain the total evaporative mass loss rate of the cloud through integration 
over $x$:
\begin{equation}
  \label{eqn:mlr_evap}
  \begin{split}
    \dot{M}_\mathrm{c,ev} &= \int_0^{\lcloud}2\pi \rcloud\mlra(x)dx \\
    &= \frac{1}{\facm}2\pi \rcloud\lcloud\mlra(0)
  \end{split}
,\end{equation}
where we introduce $\facm$ as a parameter that simplifies the integral. We approximate 
the integral by using a constant value for the mass loss rate along the cloud $\mlra(0)$ 
and apply a correction factor $\facm$ to account for actual variations along the 
cloud. Since the temperature gradient is strongest near the head ($x=0$), the conduction 
rate and the mass loss rate are also highest there. Therefore, $\facm > 1$. Appendix~\ref{sec:integrals} 
estimates that $\facm = 3.5$ under the simplified assumption that thermal conduction 
is nowhere saturated. We adopt this value $\facm = 3.5$ throughout this paper.

\subsection{The Effects of Radiative Cooling}
\label{sec:cooling}
For simulations that include thermal conduction, radiative cooling dominates over 
conductive heating over distances larger than the Field length, $\Lf$ \citep{field65}:
\bq
\Lf = \left(\frac{\kappahot \Thot}{\ncloud^2\Lambda_\mathrm{c}}\right)^{1/2}
,\eq
where $\kappahot$ and $\Thot$ are the conductive coefficient and the temperature 
of the hot ambient medium, respectively, and $\ncloud$ and $\Lambda_\mathrm{c}$ are the density 
and the cooling function of the cloud, respectively. Both analytic \citep{begelman90} 
and numerical \citep[e.g.][]{armillotta16} works suggest that clouds much larger 
than the Field length ($\rcloud \gg \Lf$) will condense as radiative cooling dominates 
and that clouds much smaller than the Field length ($\rcloud \ll \Lf$) will evaporate 
as thermal conduction dominates. Both processes need to be considered when the two 
scales are comparable to each other. It is unclear whether or not clouds will evaporate 
in this physical regime, especially if the cloud is moving relative to the ambient 
medium. Previous works that compare these two scale lengths often assume that the 
cloud is static. In this case, a temperature gradient of scale $l_\mathrm{T}$ is 
allowed to develop at the initially discontinuous interface. Since the energy exchange 
rate owing to thermal conduction scales as $T/l_\mathrm{T}^2$, conduction becomes 
less efficient as the gradient grows until $l_\mathrm{T} \sim \Lf$, where it is 
balanced by cooling. When the cloud is moving, however, the ambient flow will prevent 
such a gradient from growing thus keeping thermal conduction efficient. Therefore, 
the cloud likely still evaporates when $\rcloud \sim \Lf$.

In most of the simulations from \citetalias{bs16}, the cloud radius after the initial 
shock is smaller than $\Lf$, except for the $\chi300v1000$ simulation. Therefore, 
we assume that conduction-driven evaporation dominates over cooling-driven condensation 
in our model. However, one should be cautious about the effect of radiative cooling 
when applying our model to cosmological simulations.

Radiative cooling can also strongly affect the growth of the KHI. The evolution 
of KHI in shearing flows with cooling have been studied using numerical simulations 
that assume different geometries, e.g., 2D, 3D, slab, cylindrical, etc. Strong radiative 
cooling prevents the mixing layer at the interface from growing and penetrating 
into the cloud \citep{vietri97} and suppresses the linear growth of the KHI. However, 
whether or not cooling can enhance \citep{stone97, xu00} or suppress \citep{rossi97, 
vietri97, micono00} the long term non-linear evolution of KHI is likely sensitive 
to the details of the numerics, flow parameters, and cooling functions. Many of 
these earlier studies focus on the context of the interstellar medium (ISM), e.g., 
between proto-stellar jets and their surrounding medium of $\sim 10^4\kelvin$, where 
the physical conditions are very different from the hot halo environment.

The effect of radiative cooling has also been directly studied in cloud-crushing 
simulations. In general, efficient cooling helps compress the cloud to higher densities, 
making it more resistant to hydrodynamic instabilities \citep{klein94, armillotta16, li19}, 
but the effects are hard to quantify. This again motivates us to use a parameterized 
formula (\eqn{eqn:tkh}) to describe the KHI-driven mass loss rate of the cloud. 
Recent simulations also suggest that radiative cooling could drive thermal instabilities and cause the cloud to fragment to characteristic scales \citep{mccourt18, sparre19}, but the stripped gas from the cloud could also condense and reform cloudlets in the downstream flow under certain conditions where cooling is efficient \citep{gronke18, li19}. However, we do not model these processes in this paper.

\section{Modelling the Evolution of the Cloud}
In this section, we give a step-by-step recipe for evolving the cloud analytically 
(\sect{sec:phew}). Remember that we assume that each wind particle is a 
collection of clouds, each with a mass $\mcloud$, whose number depends on the
wind particle mass and $\mcloud$. We also summarise our main assumptions and 
approximations
and discuss the robustness of these assumptions in \sect{sec:assumptions}.

\subsection{The Analytic Model}
\label{sec:phew}
When a cloud with initial mass $\mcloud$ enters into the ambient medium at 
supersonic speed as shown in \fig{fig:cartoon}, 
we first calculate the properties related to the bow shock and the cloud shock.

\textbf{Cloud shock}. The jump conditions (Equations \ref{eqn:jc_density_ratio} and \ref{eqn:jc_temperature_ratio}) determine the post-shock pressure
\bq
\label{eqn:ana_pram}
\frac{\pii}{P_1} \sim 
\left [ \frac{2\gamma}{\gamma+1}\mach_1^2 - \frac{\gamma-1}{\gamma+1} \right] 
\etas\taus
,\eq
where $\etas$ and $\taus$ are the corrections to the jump conditions for density and 
temperature owing to thermal 
conduction (Equations \ref{eqn:jc_density} and \ref{eqn:jc_temperature}) and should 
both be 1 when thermal conduction is inefficient. The pressure across the contact 
discontinuity II is the same, i.e., $\pii = P_4$. We can solve for the post-shock 
cloud density $\rho_4$ and cloud radius $\rcloud$ under the assumption of an isothermal 
cloud shock ($\Teq = 10^4 K$):
\bq
\label{eqn:ana_rhoc}
\frac{\rho_4 \kB \Teq}{\mu m_H} = \pii
\eq
and
\bq
\label{eqn:ana_rcinit}
2\rho_4\rcloud(\pi \rcloud^2) = \mcloud.
\eq

Here, we assume that the cloud shock is nearly isotropic so that at maximum compression 
the two dimensions of the cylindrical cloud are comparable to one another, i.e., 
$\lcloud = 2\rcloud$. The cloud shock in general takes 1 to 2 cloud crushing time 
to complete.

\textbf{Confined expansion}. This only applies when thermal conduction is sufficiently strong to maintain the coherence of the cloud. When thermal conduction is weak, we proceed to calculate the mass loss rate and the deceleration of the cloud. The over-pressured cloud expands in the downstream direction at a speed $\vexp$, which we determine from \eqn{eqn:v_exp}. The length of the cloud evolves with time as $\lcloud(t) = \lcloud(t=0) + \vexp t$. We also allow the lateral dimension of the cloud $\rcloud$ to change with $\mcloud$:
\begin{equation}
\label{eqn:ana_rc}
\rcloud = \left( \frac{\mcloud}{\pi\mu m_H\ncloud\lcloud} \right)^{1/2}
,\end{equation}
where $\Ncloud$ is the total column number density along the flow direction, which 
is kept constant over time, i.e., $\Ncloud = n_4\lcloud (t=0)$. The $\rcloud$ 
calculated from \eqn{eqn:ana_rc} is consistent with the radius of the clouds in 
the numerical simulations.


\textbf{Mass loss}. The cloud loses mass owing to both the KHI (\eqn{eqn:mlr_kh}) and evaporation (\eqn{eqn:mlr_evap}). To calculate the evaporative mass loss rate per unit area at the head, i.e., $\mlra(0)$, we first determine whether or not a saturated zone exists using the criterion from \eqn{eqn:cond_zone_criteria}. If it does exist, we calculate $T_*$ by iteratively solving the equation $\sigmastar = 1$ using \eqn{eqn:cond_zone_saturation_parameter} and then find the mass loss rate using \eqn{eqn:mlra_sat}. If it does not exist, we find the mass loss rate using \eqn{eqn:mlra_class} with $T_*$ set to $\Tcloud$ in the equation.

We calculate the total mass loss rate as
\bq
\label{eqn:mlr}
\dot{M}_\mathrm{c} = \dot{M}_\mathrm{c,KH}\exp(-\rcloud/\lkh) + \dot{M}_\mathrm{c,ev}
,\eq
where $\dot{M}_\mathrm{c,KH}$ and $\dot{M}_\mathrm{c,ev}$ are mass loss rate from the KHI and evaporation alone, respectively. Since strong thermal conduction suppresses the KHI, we suppress $\dot{M}_\mathrm{c,KH}$ by a factor of $\exp(-\rcloud/\lkh)$, where $\lkh$ is determined by \eqn{eqn:lambda_kh}. Therefore, the contribution 
from KHI decreases sharply when $\rcloud \gg \lkh$ and only becomes important when 
$\rcloud \ll \lkh$.


\textbf{Deceleration}. The cloud slows down as a result of the ram pressure $\pii$. At anytime $t$, the cloud decelerates as:
\bq
\label{eqn:ana_deceleration}
\dot{v}_\mathrm{rel} = \frac{\pii\pi \rcloud^2(t)}{\mcloud(t)}
.\eq

Following the above procedures we can solve for the cloud properties 
$\rho_4(t, x)$, $\mcloud(t)$, $\vrel(t)$, $\rcloud(t)$, $\lcloud(t)$ 
at any given time by numerical integration.

\subsection{Simplifications}
\label{sec:assumptions}
Here we discuss the key simplifications in our model in the limit of strong thermal conduction. 
These simplifications are largely corroborated by the numerical simulations of \citetalias{bs16}, 
and are essential for the model to reproduce their results even qualitatively.

\textbf{Isothermal cloud}. In \citetalias{bs16}, the cloud is initially in thermal equilibrium with a temperature $\Teq \sim 10^4\kelvin$. At this temperature, radiative cooling is so efficient that during the evolution, the cloud remains nearly isothermal. Therefore, we assume that the cloud temperature $\Teq$ is invariant in our model. This also assumes that the cloud shock is isothermal, which allows the cloud to be shocked to high density. However, this assumption breaks down if the Field length is comparable to the cloud size.

In some simulations, the tail of the cloud expands so fast that during the first 
few $\tcc$ after the cloud shock parts of the cloud can be much colder than $\Teq$. 
However, the adiabatically cooled tail soon heats up and hence this deviation does 
not significantly affect the behaviour of the bulk of the cloud since most of the 
cloud mass concentrates in the dense, slowly-expanding front of the cloud.

\textbf{Constant $\qs$ parameter}. 
We use a constant value (0.9) for the $\qs$ parameter, i.e., ratio between the kinetic 
energy flux and the conductive heat flux across the bow shock, whenever thermal 
conduction dominates. In general, $\qs$ decreases from our chosen value when thermal 
conduction is sufficiently weak. However, this transition from high $\qs$ values 
(e.g., 0.9), to $\qs \sim 0$ (non-conductive) is very sharp, because the strength of 
thermal conduction is very sensitive to temperature. Therefore, deviations from this 
simplification will only affect a small range of temperatures. Moreover, since thermal 
conduction is weak in these situations, the evolution of the cloud is much less sensitive 
to the value of $\qs$ than where constant $\qs$ is a good approximation.

\textbf{Cylindrical geometry}. In our model, when thermal conduction is efficient, we let 
the cloud expand only in the downstream direction so that over time the cloud becomes 
elongated with $\lcloud \gg \rcloud$ as seen in the simulations of \citetalias{bs16}. 
The elongation helps keep the cross section of the cloud small, which keeps the cloud 
from slowing down too fast. It also results in a larger surface area between the cloud and 
the ambient flow, which makes the cloud evaporate much faster. 

\textbf{Similarity flow in the cloud}. We approximate the flow inside the cloud as an isothermal similarity flow, which parametrises the density and the pressure anywhere inside the cloud with the flow velocity only (see section \ref{sec:expansion} for details). This implies that cloud density declines logarithmically from the front to the end of the cloud, which is approximately true in \citetalias{bs16}. However, some of their simulations show that some density sub-structures emerge in the cloud later in the evolution, and that the cloud eventually breaks up into smaller aligned clumps.

\textbf{KHI suppression}. The Kelvin-Helmholtz instability, as well as other hydrodynamic instabilities that lead to the fragmentation of the cloud, are suppressed by efficient thermal conduction. This is clearly demonstrated in \citetalias{bs16}, where clouds, as long as they do not evaporate too soon, are able to maintain a coherent structure for much longer than those in the same physical conditions but without conduction \citepalias{sb15}. 

\textbf{Small vapour pressure}. We approximate that the vapour pressure is negligible compared to the post-shock thermal pressure so that the internal pressure of the cloud is balanced by thermal pressure only. Simulations indicate that at least in the front shock, the thermal pressure calculated from the conductive jump conditions are comparable to the cloud pressure except for the $\mach = 1$ cases. Inside the oblique shock (region 5), whether or not vapour pressure is important is uncertain as it is hard to compute. 

\textbf{Post-shock ambient flow}. The flow between boundary I and boundary III is a mixture of the shocked ambient gas and the evaporated material from the cloud. The flow properties here are crucial to calculating the evaporation rate from the cloud, because the conductive flux is very sensitive to the temperature gradient. To solve for the time-dependent Eulerian equations with boundary conditions at both the oblique shock front (boundary I) and the cloud surface (boundary III) is very complicated. Therefore, we simplify the problem with several approximations that are detailed in \sect{sec:conduction_zone}. Namely, we assume that the flow is continuous everywhere and can be described using Bernoulli's equations. However, when thermal conduction is too strong, e.g., in the $\chi3000v3000$ simulation where $T_1 = 3\times10^7\kelvin$, the evaporation becomes supersonic and creates shocks in the ambient flow, violating the continuity assumption. We note that the condition for supersonic evaporation is likely similar to that for vapour pressure to be dominant. In both cases, the thermal conduction must be very saturated ($\sigma_0 \gg 1$).

\textbf{No self-gravity}. The Jeans mass of the shock compressed cloud, assuming a number density of $10\,\cm^{-3}$ and a temperature of $10^4\kelvin$, is $8\times10^7\msolar$, much larger than $\mcloud$. Therefore self-gravity is almost never important in this study, unless the cloud is allowed to cool to much below the equilibrium temperature.

\section{TESTS}
\label{sec:tests}
In Figures \ref{fig:mloss}, \ref{fig:mlr_fcond} and \ref{fig:vel}, we compare the 
analytic results to simulations. For comparison, we also calculate the cloud evolution 
using a simple spherical model as described below.

\subsection{A Spherical Model}
\label{sec:model_spherical}
Semi-analytic models for clouds entrained in hot winds or clouds that travel in 
the haloes often assume the clouds are spheres with a uniform density \citep{zhang17, 
lan19}. Here, to compare with the cylindrical model, we examine whether or not a 
simpler spherical cloud model can reproduce the simulation results.

We define the properties of the cloud and the ambient medium using the same diagram as 
in \fig{fig:cartoon}. Many quantities are determined the same way as in the 
cylindrical model, except that now $\rho_4$ is constant over the cloud and $\rcloud$ 
is the radius of the sphere changing with time.

The properties of the bow shock and the cloud shock are determined by Equations 
\ref{eqn:ana_pram} and \ref{eqn:ana_rhoc}. After the cloud shock, the cloud radius 
is determined by
\bq
\label{eqn:sph_rc}
\rcloud(t) = \left( \frac{3\mcloud(t)}{4\pi\rho_4(t)}\right)^{1/3}.
\eq

At each time-step after the cloud shock, we assume that the cloud is always in pressure 
equilibrium with the post-shock ambient gas, so that $P_4 = \pii$. The cloud density 
at any time can then be derived from the pressure using \eqn{eqn:ana_rhoc}.

To calculate the evaporation rate at any given time, we use the \citetalias{cm77} 
formulation (Equations \ref{eqn:tevap_class} and \ref{eqn:tevap_sat}), which is 
derived for a spherical cloud in a static medium. Since here we only compare the 
model to conductive simulations from \citetalias{bs16}, we assume that KHI is always 
suppressed.

The deceleration of the cloud is governed by \eqn{eqn:ana_deceleration}.

\subsection{Mass Loss}
\begin{figure*}
  \includegraphics[width=1.90\columnwidth]{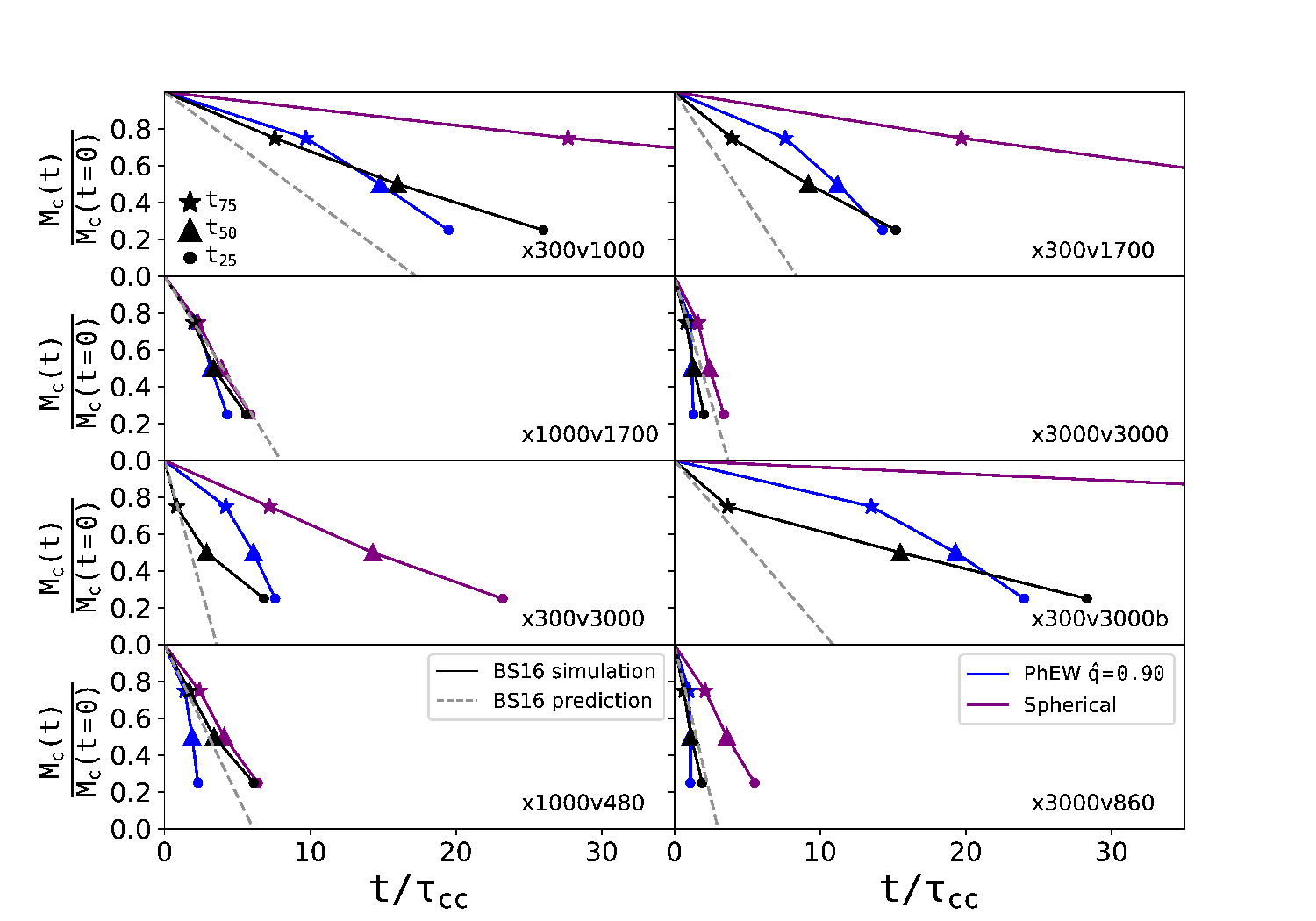}
  \centering
  \caption{Each panel shows the fraction of dense gas that remains in the cloud 
as a function of time in units of the cloud crushing time $\tcc$. Stars, triangles 
and circles indicate values at $\tseventyfive$, $\tfifty$, $\ttwentyfive$, i.e., 
when the cloud has reached 75\%, 50\% and 25\% of its original mass, respectively. 
The black symbols are results from the cloud-crushing simulations of 
\citetalias{bs16}. They provide an analytic formula (their equation 17) for the mass loss rate. We show this prediction as the grey dashed lines. 
The blue symbols are predictions from our fiducial model assuming a 
constant $\qs = 0.90$. The purple symbols are predictions from the simple spherical model (\sect{sec:model_spherical}). 
The names of the simulations are indicated in the bottom right corner of each panel.}
  \label{fig:mloss}
\end{figure*}

In \fig{fig:mloss}, we compare our model predictions of the mass evolution of the 
cloud to results from the simulations of \citetalias{bs16}, in which the cloud mass 
at any given time is defined as the total mass above a density threshold $\rho_{th} 
= \rho_1/3$, where $\rho_1$ is the original density of the cloud. This threshold 
is sufficient to capture most of the cold gas remaining in the cloud because the 
cloud shock has compressed it to a much higher density than $\rho_1$. Mass loss 
is dominated by conductive evaporation in these simulations. In our fiducial analytic 
model, the cloud mass $\mcloud(t)$ evolves with time according to \eqn{eqn:mlr_evap}. 

The spherical model presented above significantly over-estimates the 
lifetime of the cloud in most cases. The larger surface areas of the elongated 
clouds in our fiducial model play a critical role in quickly evaporating the cloud. 
The spherical model agrees with the simulations only in the two extreme cases, $\chi1000v1700$ 
and $\chi3000v3000$. In both of these cases, the shocked ambient gas is so hot ($T_5 
> 10^7\kelvin$) that some of our simplifications for the fiducial model might break down. 
First, thermal conduction is so strong that the evaporation time-scale is shorter 
than the dynamic time-scale for expansion. Second, the vapour pressure dominates 
over thermal pressure in driving the cloud shock, which in these cases compresses 
the cloud to higher densities nearly isotropically. Both of these effects tend to make the 
cloud more spherical in morphology. Therefore, the \citetalias{cm77} solution for 
spherical clouds describes the evolution of the cloud better than in the other simulations.

Our fiducial model qualitatively agrees with the simulations in all the cases shown 
here. The model over-estimates the mass loss rate for the $\chi1000v1700$ and $\chi3000v3000$ 
cases for the reasons discussed in the last paragraph. In the other simulations, the 
cloud loses mass more rapidly during the first few $\tcc$, reaching $\tseventyfive$ 
earlier than in our model, but this is because we only allow mass loss from the 
cloud after the cloud shock. During the expansion phase, our model slightly over-estimates 
the mass loss rate, e.g., in the $\chi300v3000b$ case. This is likely because of 
differences in the internal structures of the cloud at later times. In the simulations, 
density perturbations develop in the cloud with time and eventually break the cloud 
into smaller, denser clumps, but in our model we assume that the cloud always maintains 
a coherent cylindrical geometry with a logarithmic density structure, resulting 
in a larger total surface area and stronger evaporation.

In \fig{fig:mloss}, we also compare our model predictions to the analytic results derived in 
\citetalias{bs16}. They assume a constant mass loss rate from the cloud (their equation 17) 
until it completely mixes 
with the surroundings over an evaporation time-scale $t_\mathrm{evap}$ (their equation 18). The 
mass of the cloud, therefore, decreases linearly with time in their model. 
We calculate the mass loss rate according to their equations, using their fiducial parameters, 
i.e., $A = 0.01$, $T_\mathrm{evap} = 3\times10^6\kelvin$, and $\eta_\mathrm{c} = 0.5$, 
which are constrained by fitting their equations to simulation results. We show their predictions 
for the evolution of cloud mass as dashed lines in \fig{fig:mloss}. In half of these cases, 
the calculations from \citetalias{bs16} agree with our models, but in the other cases, \citetalias{bs16} 
over-estimate the mass loss rate by a factor of a few.

\begin{figure}
  \includegraphics[width=0.90\columnwidth, angle=270]{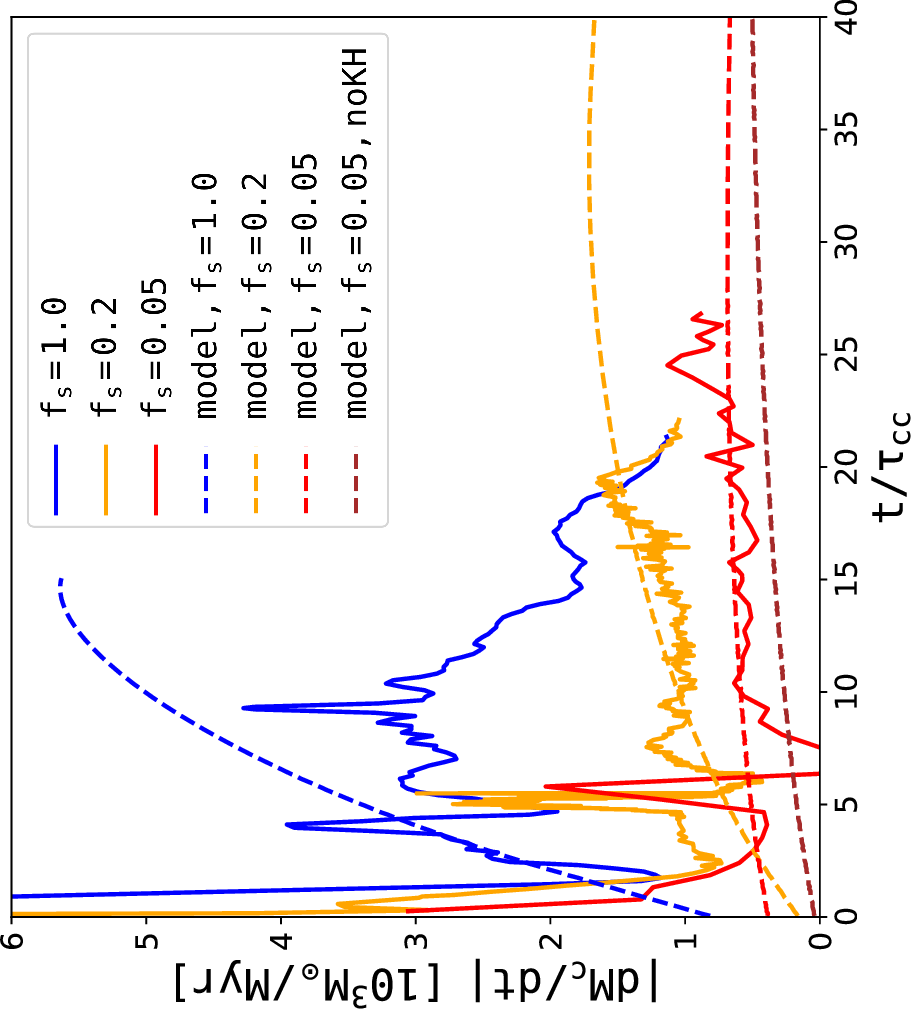}
  \centering
  \caption{The total mass loss rate as a function of time for three simulations 
with varying thermal conduction efficiencies. The $\chi300v1700$, $\chi300v1700c5$ 
and $\chi300v1700c20$ simulations are shown in blue, orange and red, respectively. 
Here we compare the solid lines from the simulations to the dashed lines from our 
model predictions. The brown dashed line shows the model prediction for the $\chi300v1700c20$ 
simulation without considering mass loss from KHI. It is the only simulation where 
KHI plays a non-negligible role while the other simulations have $\rcloud \ll \lkh$. 
The $\chi300v1700c5$ and the $\chi300v1700c20$ simulations are terminated at $\tfifty$ 
and $\tseventyfive$, respectively.}
  \label{fig:mlr_fcond}
\end{figure}

\fig{fig:mlr_fcond} demonstrates how lowering the efficiency of thermal conduction 
affects the mass loss rate. Since the cloud in the $\chi300v1700c5$ and the $\chi300v1700c20$ simulations
evaporates very slowly, we terminate those simulations at $\tfifty$ and $\tseventyfive$, 
respectively. In the first $\tcc$, there is some mass loss during the cloud shock 
in each simulation, which our model does not attempt to capture. Afterwards, our 
model agrees with the low-conduction simulations very well and also agrees with 
the $\chi300v1700$ well before $\tfifty = 9.19\tcc$. After $\tfifty$, the cloud 
in the $\chi300v1700$ simulation starts to break into clumps, shortening $\lcloud$ 
and lowering the total mass loss rate as a result. Since our model always assumes 
that the cloud is coherent, the mass loss rate from our model continues to grow 
with time as the cloud expands. In fact, for the same reason, we always over-estimate 
the late time mass loss in other simulations as well.

To first order, \eqn{eqn:mlr_evap} suggests that the mass loss rate scales linearly 
with the heat flux, so that reducing $\fS$ will also reduce $\dot{M}_\mathrm{c}$ 
by the same factor. Moreover, reducing $\fS$ changes the jump conditions at the 
bow shock, which determines the post-shock gas properties. When $\fS$ is small enough, 
however, KHI will also start to cause additional mass loss and fragmentation in 
the cloud. This is indicated by comparing the red dashed line to the brown dashed 
line in \fig{fig:mlr_fcond}. The $\chi300v1700c20$ simulation is the only one with 
$\rcloud \sim \lkh$ so that KHI causes a noticeable fraction of the mass loss.

\begin{figure*}
  \includegraphics[width=1.90\columnwidth]{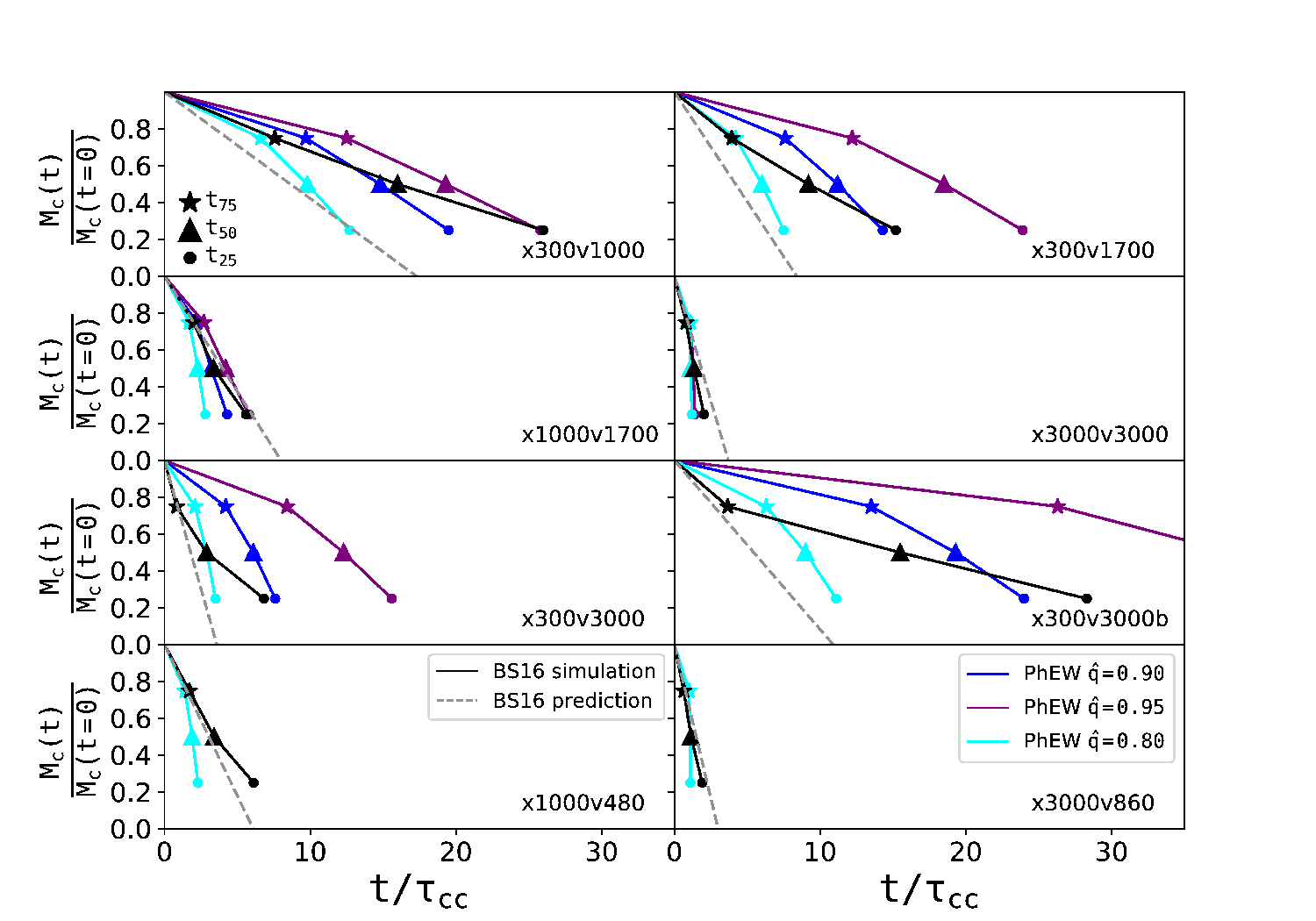}
  \centering
  \caption{Same as \fig{fig:mloss}, except that we show model predictions with three 
different $\qs$. Increasing $\qs$ results in cooler post-shock gas and less evaporation 
in general. The two simulations in the bottom panels are not affected by changing 
$\qs$ because $\mach_1 = 1$ in these simulations.}
  \label{fig:mloss_qs}
\end{figure*}

\fig{fig:mloss_qs} shows how sensitive the mass loss rate is to the $\qs$ parameter. 
At a constant Mach number, increasing $\qs$ reduces the post-shock temperature and 
increases the post-shock density (\fig{fig:qs}). As a net effect, evaporation is 
less efficient with larger $\qs$ as thermal conduction primarily depends on the 
temperature. Even though we always assume a constant $\qs$ in our model, it actually 
evolves with time. The broadening of the front shock and conduction between the 
shock and the cloud tends to increase $\qs$, making the shock more isothermal over 
time. However, we do not attempt to include this behaviour in our model, as we consider 
the model sufficiently accurate for our purposes.


\subsection{Velocity Evolution}

\begin{figure*}
  \includegraphics[width=1.90\columnwidth]{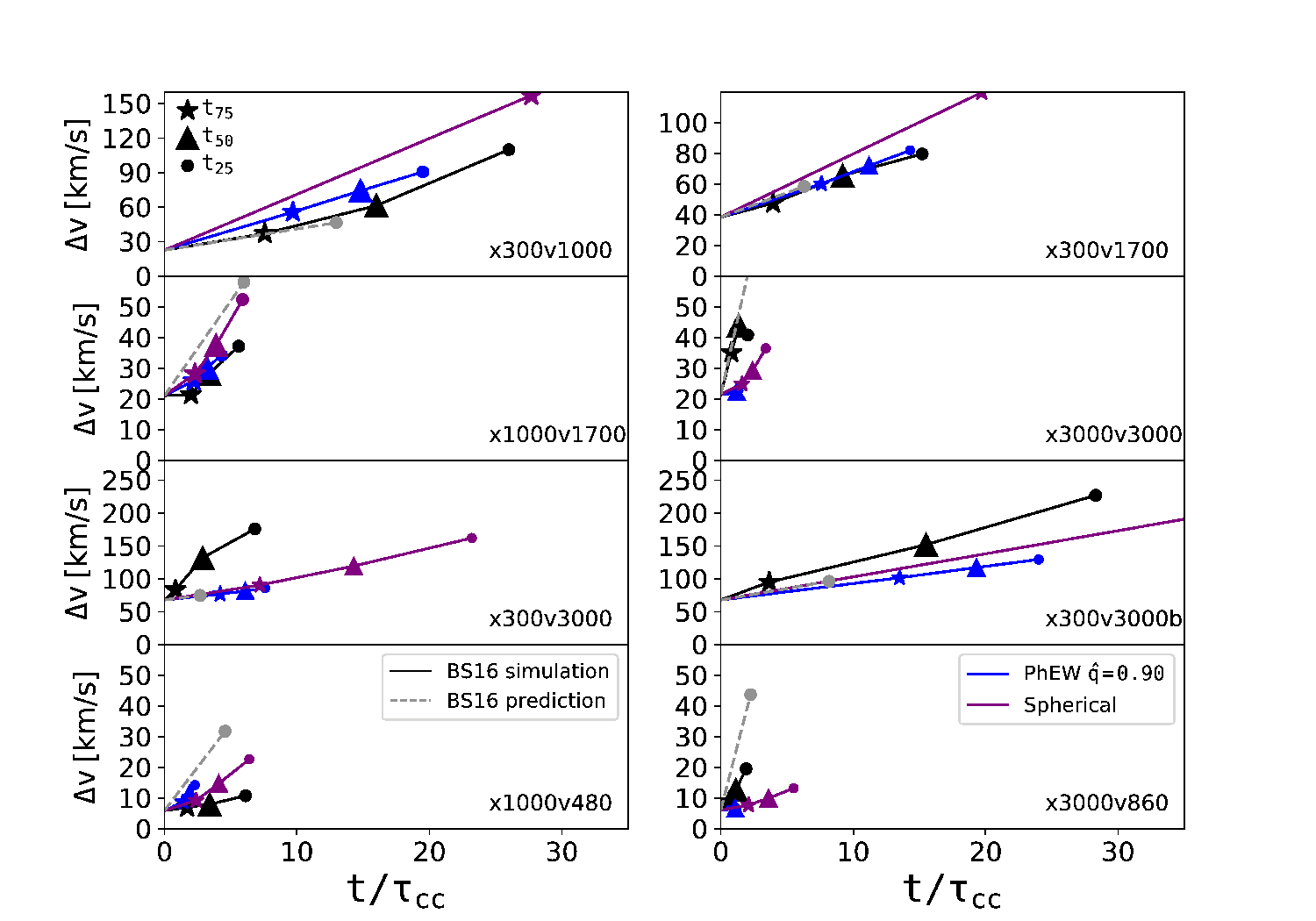}
  \centering
  \caption{The evolution of the velocity of the cloud as a function of time. In 
each panel, we compare model predictions $\Delta_v(t) + \vshock/3$ to the cloud 
speed measured from simulations. Stars, triangles and circles indicate values at 
$\tseventyfive$, $\tfifty$, $\ttwentyfive$ as in \fig{fig:mloss}. The predictions 
from our fiducial cylindrical model with $\qs = 0.90$ and the spherical model are 
shown in blue and purple, respectively. Black symbols show results from the 
simulations \citepalias{bs16}. Grey dashed lines show their analytic predictions for 
velocity evolution (their equation 23).}
  \label{fig:vel}
\end{figure*}

\fig{fig:vel} shows how the cloud's speed evolves with time. We define $\Delta v(t)$ 
as the difference between the average velocity of the cloud at any time $t$ after 
the cloud shock and the cloud velocity immediately after the cloud shock. In our 
models, $\Delta v(t)$ is governed solely by \eqn{eqn:ana_deceleration}, with $\Delta 
v(t=0) = 0$ right after the cloud shock by definition.

In the simulations, the cloud gains momentum from the cloud shock. To make fair 
comparisons between the model predicted $\Delta v$ and the cloud speed measured 
from the simulations, we calculate how much velocity the cloud gains during the 
cloud shock and add it to $\Delta v$. As an approximation, we set this initial velocity 
to $(\pi/8)\vshock$, where $\vshock$ is the shock velocity calculated by assuming 
a pressure-driven plane-parallel cloud shock. The factor $\pi/8$ comes from the 
fact that the cloud shock is not exactly plane-parallel to the cloud. Instead, the 
front half of the cloud is compressed by shocks from all sides that ultimately converge. 
We calculate the net momentum that the cloud gains from the cloud shock in the direction 
of the flow as
\bq
\frac{1}{2}(2\pi\int_0^{\rcloud} \int_{0}^{\pi} \rho_3 \vshock\sin\theta r^2\sin\theta drd\theta) = \frac{\pi}{8}\mcloud\vshock
,\eq
where $\theta$ is the angle between the radial direction of the cloud and the polar 
direction, which is the direction that is perpendicular to the flow. The constant 
factor $1/2$ takes into account the fact that only the front half of the cloud gains 
momentum from shocks.

Despite the uncertainties in the systematic offset, the velocity evolution from 
our models agrees very well with the simulation results. The slope, which corresponds 
to the deceleration rate, is well reproduced for most cases. The success of modelling 
the deceleration relies on correctly calculating the ram pressure $\pram$ and the 
cloud radius $\rcloud$, according to \eqn{eqn:ana_deceleration}. The ram pressure 
is robustly determined by the shock jump condition and is much less sensitive to 
the choice of $\qs$ than density or temperature. Therefore, correctly evolving $\rcloud$, 
and thus the cross-section for ram pressure, is key to predicting the velocity evolution. 
It is crucial that we calculate $\rcloud$ assuming cylindrical geometry and allow 
it to change only with $\mcloud$ according to \eqn{eqn:ana_rc}. The spherical models 
slightly over-estimate the deceleration rate in most cases because of their relatively 
larger $\rcloud$, according to \eqn{eqn:sph_rc}.

\citetalias{bs16} also calculate the velocity evolution of the clouds (their equation 22 and 23). 
We show their results in dashed lines in \fig{fig:vel}. Their predictions for the cloud velocities 
are very similar to ours and agree with simulations equally well, even though their derivation 
for the velocities is very different from ours. As discussed above, this agreement between our 
calculations further indicates that the velocity evolution of the cloud depends 
critically on a few quantities such as $\rcloud$ and $\pram$ that can be robustly computed. 

\section{Summary and Discussion}
\label{sec:discussion}
Hydrodynamic simulations of galaxy formation often employ sub-grid kinetic wind 
models to model feedback from star forming galaxies, however, none of the current 
simulations robustly evolve the outflowing wind material after they leave their 
host galaxies and enter into the CGM/IGM. In this paper, we propose an analytic 
model (Physically Evolved Winds; PhEW) that calculates how cold clouds that are 
launched with galactic winds evolve and propagate in such environments. We develop 
our analytic model based on findings from high resolution cloud-crushing simulations 
with \citepalias{sb15} or without including isotropic thermal conduction \citepalias{bs16} 
that simulate cold dense clouds travelling supersonically through a hot ambient medium. 

These simulations suggest that thermal conduction plays a critical role in cloud 
evolution. \citetalias{bs16} shows that strong thermal conduction changes the shock 
jump conditions, suppresses KHI, confines the cloud into a cylindrical geometry, 
and evaporates the cloud. Therefore, we build our model in two separate scenarios, 
depending on whether or not thermal conduction dominates. When thermal conduction 
is insignificant, our model predicts mass loss rates according to the empirical scaling 
relations from the non-conductive simulations of \citetalias{sb15}. Using these 
results for guidance, we self-consistently solve for the properties of the bow shock, 
the cloud shock and the evolution of the cloud. Since the strength of thermal conduction 
is very sensitive to temperature, real wind-CGM interactions in the Universe very 
likely fall into either of these scenarios. Nevertheless, we use a continuous but 
sharp transition from KHI-dominated mass loss to evaporation-dominated mass loss.

The PhEW model in thermal conduction dominated scenarios is able to predict the 
mass loss rate and the deceleration rate of the cloud at any time. These predictions 
agree with simulation results except for systems where thermal conduction is very 
saturated. We also find that a model that assumes that the clouds are spheres with 
uniform density significantly under-estimates the mass loss rate unless the evaporation 
timescale is comparable to $\tcc$.

In addition to the simulations from \citetalias{bs16}, we performed two simulations 
with reduced thermal conduction efficiency ($1/5$ and $1/20$ of the Spitzer rate) 
to $\tseventyfive$. We find that even with much weaker thermal conduction, the KHI 
is still suppressed for very long times, consistent with the findings of \citet{marcolini05}, 
where the cloud does not undergo any significant fragmentation for $\fS = 1/25$. 
The clouds in these simulations survive much longer because of their lower conductive 
evaporation rate. In the PhEW model, the KHI is nearly completely suppressed when 
$\fS = 1/5$ and is only partially suppressed when $\fS = 1/20$. Despite this difference, 
the PhEW model reproduces the mass loss rate of both clouds very well.

Many problems in galaxy formation struggle to have cold clouds survive sufficiently 
long in a hot medium. For example, entrainment of cold gas in supernova remnants 
has been proposed as a mechanism to generate galactic winds, but it is often found 
that the clouds disrupt too fast to be accelerated to wind velocities. Even after 
they are able to leave the galaxy, their subsequent evolution in the hot CGM is 
significantly limited by how fast they disintegrate. In \citetalias{bs16}, most 
clouds evaporate on a few $\tcc$, or a few Myrs, a timescale too short to be important 
for galaxy formation. Even the cloud that survives the longest can travel no more 
than $50\ \mathrm{kpc}$, a distance that is much shorter than the virial radius 
of massive, hot haloes. Furthermore, the initial mass of the cloud in their simulations is $6.7\times10^4\msolar$, 
which is likely much larger than an average cloud in the CGM. Since smaller clouds 
evaporate faster under the same physical conditions, the cloud survival problem 
becomes even more severe than that suggested by the \citetalias{bs16} simulations.

Our findings on the effects of lowering thermal conduction efficiency suggest that 
one may significantly lengthen the lifetime of clouds by keeping the thermal conduction 
very weak yet still strong enough to suppress hydrodynamic instabilities and keep the cloud 
structure coherent. For example, suppressing thermal conduction by a factor of 10 
will in principle help the cloud survive nearly 10 times longer and hence travel 
much further into the galactic halo.

Neither the \citetalias{bs16} simulations nor our models explicitly include magnetic 
fields, even though an important consequence of adding magnetic field is suppression 
of thermal conduction. Strong magnetic fields are known to also suppress hydrodynamic 
instabilities and significantly affect the geometry and lifetimes of clouds 
\citep{maclow94, orlando08, mccourt15}, though \citet{cottle20} suggest that 
magnetic draping does not significantly enhance cloud lifetime. 
Even a very weak magnetic field as probable in the CGM could strongly 
affect cloud evolution depending on the alignment between the flow and the field 
\citep{li19, cottle20}. However, our understanding of the properties and the effects of magnetic 
fields in the CGM is still very poor. Even though we do not explicitly model 
a magnetic field, we may capture its effects by varying the parameters $\fkh$ and 
$\fS$, which are, in reality, affected by the magnetic field.

It is straightforward to implement the PhEW model into hydrodynamic simulations 
of galaxy formation that employ kinetic feedback. In simulations that use a particle-based 
hydrodynamic method, e.g., smoothed particle hydrodynamic (SPH) simulations \citep{springel10b}, 
a common practice of modelling galactic winds is by statistically ejecting gas particles 
from galaxies \citep{springel03, oppenheimer06, huang20a}. 
The wind algorithm in each simulation determines the initial velocity of the ejected 
particles (wind particles) often as a function of their host galaxy properties. 
In some simulations, the wind particles temporarily decouple from the other 
SPH particles hydrodynamically after launch but soon recouple to the hydrodynamics
when the re-coupling criteria are satisfied. After re-coupling, their evolution is 
again governed by the SPH equations as for a normal gas particle.

In the PhEW model, one would launch the wind particle and let it evolve as before 
during the decoupling phase. Once it meets the original re-coupling criteria, one 
would start evolving it as a PhEW particle instead of letting it recouple. One could 
consider a PhEW particle of mass $\midx$ as a collection of $\Nidx$ identical cold 
clouds, each of them having an initial mass $\mcloud$. The cloud mass $\mcloud$ is 
a free parameter of the model but by mass conservation, $\midx = \Nidx\mcloud$. 

The choice of $\mcloud$ affects both the velocity evolution of the cloud and the mass loss rate. Under pressure equilibrium, $\rcloud$ scales with $\mcloud^{1/3}$ so that the deceleration rate scales as $\dot{v}_\mathrm{rel} \propto \mcloud^{-1/3}$ (\eqn{eqn:ana_deceleration}). When the KHI dominates the mass loss, $\tkh \propto \tcc \propto \mcloud^{1/3}$. When evaporation dominates the mass loss, the evaporation time-scale $\tau_\mathrm{ev} \propto \mcloud/\rcloud \sim \mcloud^{2/3}$. Therefore, increasing the cloud mass $\mcloud$ helps clouds survive longer. Together with $\fkh$ and $\fS$, these parameters control the evolution of PhEW particles in cosmological simulations.

To apply our PhEW model (\sect{sec:phew}) to the clouds, one would first evaluate 
the density $\rho_1$ and temperature $T_1$ of their surroundings and the relative 
velocity $\vrel$. In SPH simulations, this is conveniently done by performing a 
kernel weighted average over the neighbouring SPH particles. We would choose a time-step 
for the PhEW particle to that required for accurate integration. 
At each time-step, one would calculate the amount of mass (along with the metals), 
momentum, and energy lost since the last timestep and deposit it into the neighbouring 
SPH particles in a kernel weighted fashion. At the same time, one would reduce the 
mass and the velocity of the PhEW particle accordingly.

As a PhEW particle travels away from the galaxy into the less dense CGM/IGM, 
it will gradually expand in the radial direction and could heat up as well. 
These long-term behaviours are not modelled in the analytic model 
presented above but would need to be captured in cosmological simulations. 
In practice, one would allow the cloud radius to adjust with the ram 
pressure $\pram$ in the simulation and maintain pressure balance at the head of 
the cloud, i.e., $\pram = \ncloud\kB\Tcloud$. One would obtain the cloud radius 
under pressure equilibrium at any time using \eqn{eqn:ana_rc}:
\bq
\label{eqn:rcloud_peq}
R_\mathrm{c,peq} = \left( \frac{\gamma\mcloud}{\pi\pram\lcloud} \right)^{1/2}\cscloud.
\eq

At each time-step $\Delta t$, one would let the cloud radius adjust on a sound-crossing 
time-scale, i.e., $\tsc \equiv \rcloud/\cscloud$:
\bq
\rcloud(t + \Delta t) = R_\mathrm{c,peq} + (\rcloud - R_\mathrm{c,peq})\exp\left(-\frac{\Delta t}{\tsc} \right).
\eq

The work done by the cloud in $\Delta t$ during expansion is approximately:
\bq
W_\mathrm{exp} = 2\pi\pram\cscloud\left [\rcloud + 2(R_\mathrm{c,peq} - \rcloud)\frac{\Delta 
t}{\tsc}\right]\Delta t
,\eq
which, along with the cooling and heating rate of the cloud, determines how the 
internal energy of the cloud changes over time. 

A PhEW particle may eventually recouple if either of the following happens. 
First, it has lost over 90\% of its original 
mass. In this case one would remove the particle from simulation and deposit its 
remaining mass and momentum in the neighbouring particles. 
Second, the clouds become similar enough to the ambient medium, 
i.e., $\rho_4 \sim \rho_1$, $T_4 \sim T_1$ and $\vrel < c_1$\footnote{This is not a necessary criterion. Instead one could let the particles remain as PhEW particles. In our test simulations with PhEW, we find that most PhEW particles get destroyed by mass loss before they satisfy this recoupling criteria.}. 
Third, the particle crosses a galaxy in its path. 
In this case, which can happen in a cosmological 
simulation, the physics of PhEW would break down so we let the particle 
recouple and become a normal gas particle. We will describe the mathematical
details of this implementation in future work.

Similarly, one can combine the PhEW model with grid-based simulations. For example, 
the Illustris TNG simulations \citep{vogelsberger13, pillepich18a} model galactic 
winds by temporarily turning a cell into a particle that decouples from hydrodynamics 
until re-coupling. To apply the PhEW model to the wind particle, one would first 
track the cell where the particle is located at each time-step. Then we could use 
the cell properties as the ambient and exchange mass as well as other conserved 
quantities between the particle and the cell. Finally one would recouple the particle 
to the grid similarly as in the SPH implementation.

In summary, we developed an analytic model, PhEW, that calculates the evolution of individual clouds over a wide range of physical conditions that reproduces very high resolution simulations of individual clouds. This model can be implemented into hydrodynamic simulations of galaxy formation and will provide a more robust way of evolving cold galactic outflows in galactic haloes of various properties. The PhEW model explicitly models physical processes that occur at gas interfaces such as bow shocks, hydrodynamic instabilities, fluid mixing and thermal conduction. The PhEW model has a few parameters such as the mass of individual clouds, the Kelvin-Helmholtz coefficient $\fkh$ and the thermal conduction coefficient $\fS$ that affect the properties and the evolution of the clouds. Including these under-resolved and often neglected processes in galaxy formation simulations will be a crucial step towards a more realistic and controlled interpretation of the observations of multi-phase gas sub-structures in the circumgalactic medium within the framework of galaxy formation and evolution. We will present the results of including this model in a \textsc{GIZMO} \citep{gizmo} based cosmological simulation \citep{simba} in a future paper.

\section*{Acknowledgements}
We thank Prof. Todd Thompson and Dr. Nir Mandelker for helpful discussions. We thank Andrew Benson and Juna Kollmeier for providing computational resources at the Carnegie Institution for Science. We acknowledge support by NSF grant AST-1517503, NASA ATP grant 80NSSC18K1016, and HST Theory grant HST-AR-14299. DW acknowledges support of NSF grant AST-1909841.

\section*{Data availability}
The data underlying this article will be shared on reasonable request to the corresponding author.

\bibliography{references}

\appendix
\section{Nomenclature}
We list the main variables used in this paper here.\\
\\
\textbf{Flow properties in the cloud-crushing problem.}
\\
$\rho_1$, $T_1$, $P_1$ - density, temperature and pressure of the pre-shock ambient flow\\
$\rho_3$, $T_3$, $P_3$ - density, temperature and pressure of the cloud before the cloud shock\\
$\chi_0$ - initial density ratio between the cloud and the ambient medium\\
$\vrel$ - relative velocity between the cloud and the ambient medium\\
$\csone$ - sound speed of the pre-shock ambient flow\\
$\mach_1$ - Mach number of the ambient flow relative to the cloud\\
$\mcloud$ - cloud mass\\
$x$ - coordinate in the cloud along the long axis, with x = 0 at the cloud head\\
$\chi$ - density ratio between the cloud and the ambient medium\\
$\ncloud$, $n_4$ - hydrogen number density of the cloud\\
$\rhocloud$, $\rho_4$ - cloud density at the cloud head, i.e., short for $\rho_4(0)$\\
$\pcloud$, $P_4$ - internal pressure at the cloud head, i.e., short for $P_4(0)$\\
$\Tcloud$, $T_4$ - cloud temperature\\
$\Teq$ - cloud temperature at thermal equilibrium between radiative cooling and heating\\
$\cscloud$ - sound speed inside the cloud\\
$\rcloud$ - cloud radius perpendicular to direction of motion\\
$R_\mathrm{c,peq}$ - cloud radius under pressure equilibrium\\
$\lcloud$ - cloud length along direction of motion\\
$\pram$ - ram pressure ahead of the cloud\\
$\pii$ - pressure at the contact point II, which is equal to the ram pressure\\
$\pevap$ - vapour pressure owing to evaporation\\
$\vexp$ - expansion velocity of the cloud\\
$\rii$ - radius of the streamline that is arbitrarily chosen as the outer boundary of the conduction zone in the post-shock flow\\
$\vii$ - velocity along the streamline\\
$\nii$ - hydrogen number density along the streamline\\
$\rhoii$ - density along the streamline\\
$\Tii$ - temperature along the streamline\\
$\cii$ - sound speed along the streamline\\
\\
\textbf{Properties of the bow shock and the cloud shock.}
\\
$\vshock$ - velocity of the cloud shock\\
$\rho_\mathrm{a}$ - density of the pre-shock gas\\
$T_\mathrm{a}$ - temperature of the pre-shock gas\\
$\rho_\mathrm{ps}$ - density of the post-shock gas\\
$T_\mathrm{ps}$ - temperature of the post-shock gas\\
$\etas$ - correction factor for the density jump across a conductive shock\\
$\taus$ - correction factor for the temperature jump across a conductive shock\\
\\
\textbf{Thermal conduction.}
\\
$\nhot$ - hydrogen number density in the hot gas in a two-phase medium\\
$\rhohot$ - gas density in the hot gas in a two-phase medium\\
$\Thot$ - temperature of the hot gas in a two-phase medium\\
$\chot$ - sound speed of the hot gas in a two-phase medium\\
$\machhot$ - Mach number of the hot gas in a two-phase medium\\
$\mfp$ - mean free path of electrons in the hot medium\\
$\qclass$ - classical heat flux from thermal conduction\\
$\qsat$ - saturated heat flux from thermal conduction\\
$\sigma_0$ - conductive coefficient, defined as the ratio between the classical and the saturated heat flux\\
$\sigmacloud$ - conductive coefficient at the cloud surface\\
$\tau_\mathrm{ev,class}$ - time-scale for classical evaporation from \citetalias{cm77}\\
$\tau_\mathrm{ev,sat}$ - time-scale for saturated evaporation from \citetalias{cm77}\\
\\
\textbf{The conduction zone.}
\\
$\machsat$ - Mach number in the saturated zone \\
$r_*$ - radius of the transition point where thermal conduction saturates\\
$n_*$ - hydrogen number density at the transition point\\
$\rho_*$ - density at the transition point\\
$T_*$ - temperature at the transition point\\
$\qstar$ - Heat flux at the transition point\\
\\
\textbf{Mass loss rates.}
\\
$\mlra$ - mass loss rate per unit area owing to conductive evaporation on the cloud surface\\
$\dot{M}_\mathrm{c,KH}$ - total mass loss rate from KHI\\
$\dot{M}_\mathrm{c,ev}$ - total mass loss rate from conductive evaporation\\
$\dot{M}_\mathrm{c}$ - total mass loss rate of the cloud from both KHI and conductive evaporation\\
\\
\textbf{Various scales.}
\\
$\tsc$ - sound crossing time-scale\\
$\tcc$ - cloud-crushing time-scale\\
$\tdiff$ - diffusion time-scale owing to thermal conduction\\
$\tmix$ - mixing time-scale owing to Kelvin-Helmholtz instability\\
$\tkh$ - Kelvin-Helmholtz time-scale\\
$\lkh$ - Kelvin-Helmholtz scale. Perturbations on scales below it are suppressed by thermal conduction\\
$\Lf$ - Field length\\
\\
\textbf{Parameters and fixed-value factors.}
\\
$\fS$ - parameter that determines the efficiency of thermal conduction. $\fS = 1$ corresponds to conduction at the Spitzer value.\\
$\fkh$ - parameter that determines the general growth of Kelvin-Helmholtz instability mass loss\\
$\qs$ - the ratio between the heat flux and the kinetic energy flow across a conductive shock front, approximated as 0.90 in this paper\\
$\fram$ - factor that affects the ram pressure according to \eqn{eqn:P_ram}, approximated as 0.5 in this paper\\
$\fr$ - factor defined as $\fr \equiv \ln(\rii/\rcloud)$, approximated as 1.0 in this paper\\
$\facm$ - factor that affects the total evaporation rate according to \eqn{eqn:mlr_evap}, approximated as 3.5 in this paper\\

\section{Modified Shock Jump Conditions}
\label{sec:shock_jump_conditions}
The Rankine-Hugoniot jump conditions relate post-shock gas properties to the pre-shock 
gas properties across an adiabatic, non-conductive plane-parallel shock. When thermal 
conduction is efficient, the shock front will be smoothed by the enthalpy flow in 
the upstream direction. The jump conditions can be obtained by considering that 
fluid quantities are conserved across the shock:
\begin{equation}
  \label{eqn:jp_equations_1}
  \rho_1v_1 = \rho_2v_2,
\end{equation}
\begin{equation}
  \label{eqn:jp_equations_2}
  \rho_1v_1^2 + P_1 = \rho_2v_2^2 + P_2,
\end{equation}
and
\begin{equation}
  \label{eqn:jp_equations_3}
  \frac{1}{2}\rho_1v_1^3 + \frac{5}{2}P_1v_1 =  \frac{1}{2}\rho_2v_2^3 + \frac{5}{2}P_2v_2 + q_s  
,\end{equation}
where, following the notation of \citet{borkowski89}, we introduce dimensionless 
parameters $\etas \equiv \rho_2/(4\rho_1)$, $\taus \equiv 16\kB T/(3\mu \mh v_1^2)$, 
and $\qs \equiv q_s / (\rho_1v_1^3/2)$.

Equations~\ref{eqn:jp_equations_1} to \ref{eqn:jp_equations_3} are identical to 
equations (8) to (10) in \citet{borkowski89} except for the pre-shock terms $P_1$ 
and $5P_1v_1/2$ that are ignored in their paper. Even though $P_2 \gg P_1$ in equations~\ref{eqn:jp_equations_2} 
and \ref{eqn:jp_equations_3}, $P_1v_1/P_2v_2 \sim T_1/T_2$ is not necessarily infinitesimal, 
unless the shock mach number $\mach \gg 1$. Therefore, including these two terms, 
especially the second, should largely improve the accuracy in low Mach number shocks.

Solving these equations we obtain:
\begin{equation}
\label{eqn:jc_density}
\etas = \frac{5(1+\betas) + \sqrt{9+16\qs+5\betas(5\betas-6)}}{8(1-\qs+5\betas)},
\end{equation}
and
\begin{equation}
\label{eqn:jc_temperature}
\taus = \frac{1}{2} - \frac{4\qs}{3} + \frac{1}{6}(1+\betas)\sqrt{9+16\qs+5\betas(5\betas-6)} + \frac{5}{6}\betas(\betas + 6),
\end{equation}
which immediately yields Equations \ref{eqn:jc_density_ratio} and \ref{eqn:jc_temperature_ratio} 
in the text.





\section{Integrals}
\label{sec:integrals}
To integrate Equation \ref{eqn:mlr_evap} one needs to know the mass loss rate 
per area, $\mlra$, at each point along the cloud, which in turn relies on the properties 
of the ambient flow. We relate each point along the cloud to another point on the 
streamline (noted as boundary II in Figures \ref{fig:cartoon} and \ref{fig:conduction_zone}) 
in the ambient flow. Along the streamline, fluid properties are parameterised by 
flow velocity according to Bernoulli's equations:
\begin{equation}
  \label{eqn:bernouli_density}
  \frac{\rhoii(0)}{\rhoii(x)} = \left( 1 - \frac{\gamma - 1}{2}\frac{\vii^2(x)}{\cii^2(x)}\right)^\frac{1}{\gamma-1},
\end{equation}
\begin{equation}
  \label{eqn:bernouli_pressure}
  \frac{\pii(0)}{\pii(x)} = \left( 1 - \frac{\gamma - 1}{2}\frac{\vii^2(x)}{\cii^2(x)}\right)^\frac{\gamma}{\gamma-1},
\end{equation}
and
\begin{equation}
  \label{eqn:bernouli_temperature}
  \frac{\Tii(0)}{\Tii(x)} = 1 - \frac{\gamma - 1}{2}\frac{\vii^2(x)}{\cii^2(x)}
,\end{equation}
where $\vii(x)$ and $\cii(x)$ are the flow speed and the sound speed at the point 
on the streamline that corresponds to coordinate $x$ in the cloud. The cloud head 
at $x=0$ corresponds to the stagnation point where $\vii = 0$.

To simplify the integral, we assume that
\begin{equation}
  \label{eqn:bernouli_assumption}
  \frac{dx}{d\vii^2} = \frac{x}{\vii^2}.
  \end{equation}

Also, we assume that at the tail of the cloud $x=\lcloud \gg \rcloud$, the ambient 
flow becomes identical to the unperturbed flow: $\Tii(\lcloud) = T_1$, $\cii(\lcloud) 
= c_1$.

Combining Equation \ref{eqn:bernouli_temperature} and Equation \ref{eqn:bernouli_assumption}:
\begin{equation}
  \begin{split}
    dx &= \left( \frac{\lcloud}{\vii^2(\lcloud)} \right)d\vii^2\\
    &= \left[ \frac{(\gamma - 1)\lcloud}{2c_1}\left(1 - \frac{T_1}{\Tii(0)}\right)^{-1}\right]d\vii^2\\
    &= -\lcloud\left( 1-\frac{T_1}{\Tii(0)} \right)^{-1}\frac{dT}{\Tii(0)}\\
    &= -\lcloud\frac{dT}{\Tii(0)-T_1}
  \end{split}
\end{equation}

Therefore, assuming classical conduction, the integral governing the total mass 
loss rate becomes:
\begin{equation}
  \begin{split}
    \int_0^{\lcloud} \mlra(x) dx
    &= A(T_1, \Tii(0), \rcloud)\frac{\lcloud}{\Tii(0)-T_1}\int_{T_1}^{\Tii(0)} T^{5/2}dT \\
    &\sim \frac{1}{3.5}A(T_1, \Tii(0), \rcloud)\lcloud\frac{\Tii(0)^{7/2}-T_1^{7/2}}{\Tii(0)-T_1}\\
    &= \frac{1}{3.5}\left[\frac{1-\left(\frac{T_1}{\Tii(0)}\right)^{3.5}}{1-\left(\frac{T_1}{\Tii(0)}\right)} \right]\lcloud\mlra(0)
  \end{split}
.\end{equation}

Comparing the above results to Equation \ref{eqn:mlr_evap} indicates $\facm = 3.5$ 
when $\Tii(0) \gg T_1$.

\bsp
\label{lastpage}
\end{document}